\documentclass[extra,fleqn]{gji}
\usepackage{timet}

\usepackage{color}

\usepackage{amsmath,amssymb}

\usepackage{graphicx}
\usepackage[normalem]{ulem}

\graphicspath{{Figs/}}

\newcommand{\derpar}[2]{\ensuremath{\frac{\partial #1}{\partial #2}}}

\usepackage[colorlinks=true]{hyperref}

\begin{document}
\title{Earth's Inner Core dynamics induced by the Lorentz force}
\author[M. Lasbleis, R. Deguen, P. Cardin, S. Labrosse]{M.  Lasbleis$^1$, R. Deguen$^1$, P. Cardin$^2$,
  S. Labrosse$^1$ \\$^{1}$LGL-TPE, Universit\'e Lyon 1-ENS de Lyon-CNRS, Lyon, France, $^{2}$ ISTerre, UGA-CNRS, Grenoble, France}

\maketitle

\begin{summary}
Seismic studies indicate that the Earth's inner core has a complex
structure and exhibits  a 
strong elastic anisotropy with a cylindrical symmetry. 
Among the various models which have been proposed to explain this
anisotropy,  
one class of models considers the effect of the Lorentz force
associated with the magnetic field diffused within the inner core. 
In this paper we extend previous studies and use analytical calculations
and numerical 
simulations  to predict the geometry and strength of the flow induced
by the poloidal component of the Lorentz force in a neutrally or stably stratified 
growing inner core, exploring also the effect of different types of
boundary conditions at the inner core boundary (ICB).  
Unlike previous studies, we show that the boundary condition that is most likely to produce a significant
  deformation and seismic anisotropy is impermeable, with
negligible radial flow through the boundary.
Exact analytical solutions are found in the case of a negligible
effect of buoyancy forces in the inner core (neutral stratification),
while numerical simulations are used to investigate the case of 
stable stratification. 
In this situation, the flow induced by the Lorentz force is found to
be localized in a shear layer below the ICB, which thickness depends
on the strength of the stratification, but not on the magnetic field
strength. 
We obtain scaling laws for the thickness of this layer, as well as for
the flow velocity and strain rate in this shear layer as a function of the
 control parameters, which include the magnitude of the magnetic
field, the strength of the density stratification,  the viscosity of
the inner core, and the growth rate of the inner core. 
We find that the resulting strain rate is probably too small to
produce significant texturing unless the inner core viscosity is
smaller than about $10^{12}$ Pa.s.  
\end{summary}


\section{Introduction}
\label{sec:introduction}

The existence of structures within the inner core was first discovered
by \citet{POUPINET:1983ue}, who discussed the possibility of lateral
heterogeneity from the observation of P-waves travel time anomalies. These  were then
attributed to the existence of seismic anisotropy
\citep{Morelli:1986ts,Woodhouse:2012bh}, with P-waves travelling
faster in the north-south direction than in the equatorial
plane. Since then, more complexities have been discovered in the inner
core: a slight tilt in the fast axis of the anisotropy, radial
variations of the anisotropy with a nearly isotropic upper layer, 
hemispherical variations of the thickness of the upper isotropic layer,
an innermost inner core with different properties in anisotropy or
attenuation, and anisotropic attenuation \citep[See ][ for reviews, and
references
therein]{Souriau:2003il,Tkalcic2008,Deguen:2012kx,Deuss:2014gz}. 	

The seismic anisotropy 
can be
explained either by liquid inclusions elongated in some specific
direction (shape preferred orientation, SPO) \citep{Singh:2000uk} or
by the alignment of the iron crystals forming the inner core (lattice
preferred orientation, LPO). In the case of LPO, the orientation is acquired either
during crystallization \citep[\textit{e.g.} ][]{Karato:1993vy,Bergman:1997tw,Brito:2002} or by 
texturing during deformation of the inner core. Several mechanisms
have been proposed to provide the deformation needed for texturing:
solid state convection
\citep{Jeanloz:1988wa,Weber:1992ts,Buffett:2009ka,Deguen:2011ga,Cottaar:2012hg,Deguen:2013bj},
or deformation induced by external forcing,  due to viscous adjustment
following preferential growth at the equator
\citep{Yoshida:1996p54,Yoshida:1998um,Deguen:2009ej}, or Lorentz force
\citep{Karato:1999tt,Buffett:2000tb,Buffett:2001ul}.

Thermal convection in the inner core is possible if its cooling rate,
related to its growth rate, or radiogenic heating rate is large enough
to maintain a temperature gradient steeper than the isentropic gradient. In
other words, the heat loss of the inner core must be larger than what
would be conducted down the isentrope. However, the thermal
conductivity of the core has been recently reevaluated to values
larger than 90\,W.m$^{-1}$.K$^{-1}$ at the core mantle boundary and in excess of
150\,W.m$^{-1}$.K$^{-1}$ in the inner core
\citep{deKoker:2012wh,Pozzo:2012ty,Gomi:2013jp,Pozzo:2014vb},
and this makes thermal convection in the inner core unlikely
\citep{Yukutake:1998td,Deguen:2011ga,Deguen:2013bj,Labrosse:2014hf}. Inner
core translation, that has been proposed to explain the
hemispherical dichotomy of the inner core \citep{Monnereau:2010bma}, results
from a convection instability
\citep{Alboussiere:2010p24,Deguen:2013bj,Mizzon:2013tm} and is	
therefore also difficult to sustain.

Compositional convection is possible if the partition coefficient of
light elements at the inner core boundary (ICB) decreases with time
\citep{Deguen:2011ga,Gubbins:2013ip} or if some sort of compositional
stratification develops in the outer core
\citep{Alboussiere:2010p24,Buffett:2000gb,Gubbins:2013bx,Deguen:2013bj}
so that the concentration of the liquid that crystallizes decreases
with time. However, the combination of both thermal and compositional
buoyancy does not favor convection in the inner core \citep{Labrosse:2014hf}.

The strong thermal stability of the inner core resulting from its high
thermal conductivity \citep{Labrosse:2014hf} is a barrier to any vertical
motion and other forcing mechanisms need to work against it. This
situation has already been considered in the case of deformation
induced by preferential growth in the equatorial belt \citep{Deguen:2009ej},
and has been shown to produce a layered structure. \citet{Deguen:2011ck} and \citet{Lincot:2014hy}
evaluated the predictions of anisotropy from this model and found that
although it can induce significant deformation, it is
difficult to explain the strength and geometry of the anisotropy
observed in the inner core.

In this paper, we consider another major external forcing that was
proposed, Maxwell stress. This was first proposed by
\citet{Karato:1999tt} who considered the action of the Lorentz force 
assuming the inner core to be neutrally buoyant
throughout. This situation is rather unlikely and, as discussed above,
we expect the inner core to be stably
stratified. \citet{Buffett:2000tb} have shown that in this case the
flow is confined in a thin layer at the top of the inner core, similar
to the case discussed above for a flow driven by preferential growth
at the equator. However,  
the growth of the inner core gradually buries the deformed iron
and this scenario may still produce a texture in the whole inner
core. All these previous studies considered a fixed inner core size
and infinitely fast phase change at the ICB. 
The moving
boundary brings an additional advection term in the heat balance which can
influence the dynamics. 
In the context
of inner core convection \citet{Alboussiere:2010p24} and
\citet{Deguen:2013bj} have proposed a boundary condition at the ICB that
allows for a continuous variation from perfectly permeable boundary conditions,
that was considered in previous studies, to  impermeable boundary conditions
when the timescale for phase change is large compared to that for
viscous adjustment of the topography. 

In this paper, we investigate the dynamics of a growing inner core
subject to electromagnetic forcing, and include the effects of a stable
stratification, of the growth of the inner core, and different types of boundary
conditions. We propose a systematic study of the dynamics induced by a poloidal Lorentz
force in the
inner core and develop scaling laws to estimate the strain rate of the flow. 

In Section \ref{sec:governing-equations} , we develop a set of equations taking into account  the Lorentz
force and a buoyancy force from either thermal or compositional origin. Analytical and numerical
results are presented in Section \ref{sec:flow-description}, scaling laws for
the maximum velocity and strain rate are developed in Sections \ref{sec:scal-laws-numer}
and \ref{sec:strain-rate-produced} and compared to numerical solutions.   In Section \ref{sec:discussr}, we use our results to predict the instantaneous  strain rates and
cumulative strain in the Earth's inner core due to the Lorentz force.

\section{Governing equations}
\label{sec:governing-equations}

\subsection{Effect of an imposed external magnetic field}
\label{sec:effect-an-imposed}

The magnetic field produced by dynamo action in the liquid outer core extends up to
the surface of the Earth, but also to the center-most part of the
core. 
Considering for example a flow velocity of the order of the growth
rate of the inner core gives a magnetic Reynolds number (comparing advection and
diffusion of the magnetic field)  of the
inner core of about $10^{-5}$. This shows that the magnetic field in
the inner core is only maintained by diffusion from its boundary. 

Two dynamical effects need to be
taken into account: the Lorentz force and  Joule
heating. The
Lorentz force acts directly on the momentum conservation, while  Joule heating is part
of the energy budget and modifies the temperature distribution, inducing a flow
through buoyancy forces.

In this paper, we will discuss the effect of the Lorentz force in the case
of a purely toroidal axisymmetric magnetic field with a simple mathematical form. The effect of Joule heating in the case of
a 
non growing inner core was studied by \citet{Takehiro:2010cw} and will not be
investigated further here.

The poloidal magnetic field intensity at the core mantle boundary (CMB) can be inferred from surface
observations of the field at the Earth's surface, but both poloidal and
toroidal components are poorly known deeper in the core. 
The root mean square (RMS) strength of the
field at the ICB has been estimated using  numerical simulations
to be around a few milliteslas
\citep[e.g.][]{Glatzmaier:1996uj,Christensen:2006jo}. It can be also constrained by
physical observations:  for example, \citet{Koot:2013cm} give an upper bound of 9-16mT for
the RMS field at the ICB by looking at the dissipation in the electromagnetic coupling,
while \citet{Gillet:2010p6} suggest 2-3mT from the observation of fast toroidal
oscillations in the core. \citet{Buffett:2010ky} obtains similar values from  measurements of tidal dissipation.
Numerical simulations also predict a strong azimuthal component $B_{\phi}$ at the vicinity of the
inner core, possibly one order of magnitude higher than the vertical component $B_{z}$
\citep{Glatzmaier:1996uj}, though this depends on the magnitude of inner core differential rotation. 

\citet{Buffett:2001ul} have considered the effect of the azimuthal component of the Lorentz force
resulting from the combination of the $B_{z}$ and $B_{\phi}$ components of the magnetic field. 
We will focus here on the effect of the azimuthal component of the magnetic field, for
which the associated  Lorentz force is poloidal and axisymmetric. 
The flow calculated by \citet{Buffett:2001ul} is decoupled from the flow induced by the
azimuthal component of the magnetic field, and thus the total axisymmetric flow can be obtained by
simply summing  the two flows.

One of the most intriguing feature of the Earth's magnetic field is the existence of
reversals. However, since the Lorentz force depends quadratically on the magnetic field, its direction is not modified by a reversal of the field. 
For
simplicity, we will
consider that the magnetic field is constant in time. 

The magnetic field inside the inner core is calculated by diffusing the field
from the ICB. 
The magnetic Reynolds number for the inner core being very small, $\mitbf{B}$ is not advected by the flow. 
Because the seismic observation of anisotropy is of large scale, 
and also because low-order toroidal component penetrates deeper
inside the inner core,  only the lowest order of the azimuthal
component of the magnetic field is taken into account, following the work of \citet{Karato:1999tt} and
\citet{Buffett:2000tb}.

We consider a purely toroidal axisymmetric field of degree two  in the vicinity of  the
ICB, of the form   $\left . \mitbf{B}\right| _{ICB}=B_0 \sin \theta \cos \theta
\mitbf{e}_{\phi}$ \citep{Buffett:2000tb}.  Solving $\nabla ^2 \mitbf{B}=0$,  the 
field inside the inner core is   
 \begin{equation}
   \mitbf{B}=B_{0} \frac{r^2}{r_{ic}^2} \cos \theta \sin \theta \mitbf{e}_{\phi},\label{eq:Binside}
 \end{equation}   
in spherical coordinates, which is associated to an electric current density $\mitbf{J}=\frac{1}{\mu_0} \mitbf{\nabla}\times\mitbf{B}$,
where $r_\mathrm{ic}$ is the radius of the inner core and 
$\mu_0$ is the magnetic permeability. 

The Lorentz force is a  
volume force
given by  $\mitbf{F}_L=\mitbf{J} \times \mitbf{B}$. 
The Lorentz force can be decomposed as the sum of the gradient of a magnetic
pressure and a non-potential part as $\mitbf{F}_L=-\mitbf{\nabla}P_m +\mitbf{f}_L$, which
is a unique Helmholtz decomposition for $\mitbf{\nabla}\cdot
\mitbf{f}_L=0$. 
 With the magnetic field as 
defined in Eq.~\eqref{eq:Binside}, we find that $P_m$ and $\mitbf{f}_L$ are given by
\begin{equation}
\label{eq:Pm}
P_m = \frac{1}{7} \frac{B_{0}^{2}}{\mu_{0}}\frac{r^{4}}{r_{ic}^{4}} \left( \frac{3}{2} \cos^{2}\theta + \frac{1}{5}  \right)
\end{equation}
and
\begin{multline}
\label{eq:fL}
\mitbf{f}_L= \frac{B_{0}^{2}}{\mu_{0} r_{ic}}\frac{r^{3}}{r_{ic}^{3}}  \left[\left(3 \cos^{4}\theta - \frac{15}{7}\cos^{2}\theta + \frac{4}{35} \right)\mitbf{e}_r \right.\\
\left. + \cos \theta \sin \theta \left(\frac{4}{7} - 3\cos^{2}\theta \right)\mitbf{e}_\theta \right].
\end{multline}

\begin{figure}
  \centering
  \includegraphics{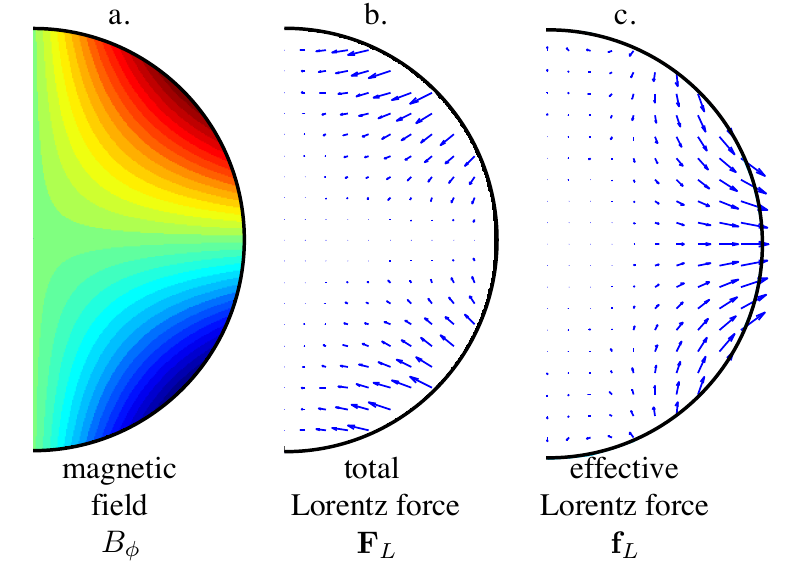}
  \caption{Meridional cross sections showing the intensity of the magnetic field (a), the Lorentz
    force $\mitbf{F}_L$ (b) and  the non-potential part of the 
    Lorentz force $\mitbf{f}_L$ as defined in equation~(\ref{eq:fL})
    (c). }
  \label{fig:Forces}
\end{figure}

The potential part of the Lorentz force can only promote a new equilibrium state but no
persisting flow. We are thus only interested in the non potential part of the Lorentz
force, shown in Fig.~\ref{fig:Forces}. Eq.~\eqref{eq:fL} provides a characteristic scale for the force as $B_0^2/\mu_0 r_{ic}$.

\citet{Karato:1999tt} investigated the effect of the Maxwell stress by applying a given
normal stress on the inner core boundary. This is different from our study, where, as in \cite{Buffett:2000tb}, we
consider a volumetric forcing, as shown on Fig.~\ref{fig:Forces}, and not a forcing on the surface of the inner core.

\subsection{Conservation equations}
\label{sec:cons-equat}

\subsubsection{Conservation of mass, momentum and energy}
\label{sec:cons-mass-moment}

We consider 
an incompressible fluid in a spherical domain, with a newtonian rheology of uniform
viscosity $\eta$,  neglecting inertia. Volume forces considered here are the
buoyancy forces, with density variations due to temperature or compositional variations, and
the Lorentz force as defined above. 

The equations of continuity and conservation of momentum are written as 
\begin{equation}
  \mitbf{\nabla}\cdot  \mitbf{u}=0,  \label{eq:setofequations1}
\end{equation}
\begin{equation}
\mitbf{0}=-\mitbf{\nabla} p'+\Delta \rho\, \mitbf{g} +\eta \nabla  ^2 \mitbf{u}+\mitbf{f}_L,
\label{eq:setofequations2}
\end{equation}
where $\mitbf{u}$ is the velocity, $p'$ the dynamic pressure that also includes the magnetic pressure,
$\Delta \rho$ the density difference compared to the reference density
profile,  and $\mitbf{g}=g_{ic}r/r_{ic} \mitbf{e}_r$ the acceleration
of gravity with $g_{ic}$ the acceleration of gravity at $r=r_{ic}$.

The density depends on both the temperature $T$ and the light element
concentration $c$. 
We define a potential temperature as $\Theta=T-T_s(r,t)$, with $T_s(r,t)$ the
isentropic temperature profile anchored to the liquidus at the
ICB, and introduce a potential composition $C=c-c^s_{ic}(t)$, where
$c^s_{ic}(t)$ is the composition of the solid at the ICB. 
We will consider separately the effects of composition and temperature, but both
can induce a density  stratification, which is quantified through a variation of density $\Delta \rho $
which is either $\rho \alpha_T \Theta$ or $\rho \alpha_C C$, where
$\rho$ is the reference density, and $\alpha_T$ and $\alpha_C$ the
coefficients of thermal and compositional expansion, respectively. Because the potential temperature and composition are
solutions of mathematically similar equations, we will use a quantity $\chi$
 which is either the potential temperature $\Theta $ or composition $C$.
 In this paper, quantities that apply
 for both cases will have no subscript, whereas we will use $_T$ for quantities referring
 to the thermal stratification, and $_C$ for compositional stratification.

The momentum conservation equation (\ref{eq:setofequations2}) is thus written as
\begin{equation}
  \label{eq:momentumconservation}
  \mitbf{0}=-\mitbf{\nabla} p'+\alpha \rho\chi\, g_{ic}\frac{r}{r_{ic}} \mitbf{e}_r +\eta \nabla  ^2 \mitbf{u}+\mitbf{f}_L.
\end{equation} 
The equations for the evolution of the potential temperature (energy
conservation) and of light element concentration (solute conservation)
have a common form, which will be written as 
\begin{equation}
\derpar{\chi}{t}+\mitbf{u}\cdot  \mitbf{\nabla} \chi=\kappa \nabla ^2 \chi +S(t),   \label{eq:setofequations3}
\end{equation}
where $\kappa$ is the diffusivity of either heat ($\kappa_T$) or composition ($\kappa_C$) and $S$
a source term 
given by
\begin{equation}
  \label{eq:St}
  S_T(t)=\kappa_T \nabla^2T_s-\derpar{T_s}{t}
\end{equation}
and
\begin{equation}
  \label{eq:Sct}
  S_C(t)=-\derpar{c_{ic}^s}{t}.
\end{equation}

As discussed in  \citet{Deguen:2011ga},  the inner core is
stably stratified when the source term $S(t)$ is negative, and no convective instability can develop. In this paper, we will
focus on this case, with either $S_T(t)$ or $S_C(t)$ negative.

\subsubsection{Growth of the inner core}
\label{sec:growth-inner-core}

 To take into
account the growth of the inner core, we use a front fixing approach to solve the moving
boundary problem \citep{Crank:1984uf} by  scaling lengths with the
inner core radius $r_{ic}(t)$ at time $t$. We define a new coordinate
system with $\tilde{r}=r/r_{ic}(t)$. This modifies slightly the spatial derivatives by bringing a
factor
$1/r_{ic}(t)$ to radial derivatives, but also adds a  radial advection term in the equations where the time
derivative is present. In the new coordinate
system, we obtain 
\begin{equation}
  \label{eq:newcoordinates}
  \left . \derpar{}{t}\right | _{\tilde{r}}
= \left
    . \derpar{}{t}\right | _{r}+\tilde{r}\frac{u_{ic} (t)}{r_{ic}(t)}  \left . \derpar{}{\tilde{r}}\right | _{t},
\end{equation}
where $u_{ic}(t)=dr_{ic}/dt$ is the instantaneous growth rate of the inner core. 
Eq.~(\ref{eq:setofequations3}) becomes
\begin{equation}
  \label{eq:temperature_r'x}
  \derpar{\chi}{t}+\frac{1}{r_{ic}(t)}(\mitbf{u}-\tilde{r}\,u_{ic}(t)\mitbf{e}_r)\cdot  \mitbf{\nabla}\chi=\frac{\kappa}{r^2_{ic}(t)} \nabla ^2 \chi +S(t), 
\end{equation}
where $\cdot \mitbf{\nabla}$ and $\nabla ^2$ are now spatial derivative operators in the new coordinate system
$(\tilde{r},\theta,\phi)$, with $\theta$  and $\phi$ the  colatitude and longitude.

\subsection{Dimensionless equations and parameters}
\label{sec:dimens-equat-param}

\subsubsection{Definition of the dimensionless quantities}
\label{sec:defin-dimens-quant}

The set of equations (\ref{eq:setofequations1}), (\ref{eq:momentumconservation}),
\eqref{eq:setofequations3} is now made dimensionless, using 
$r_{ic}(t)$,  the age of the inner core $\tau_{ic}$, $\kappa /r_{ic}(t)$, $\eta \kappa /r_{ic}^2 (t)$ and
$\Delta \rho_\chi $ as characteristic scales for length, time, velocities, pressure, and
density variations. 
The density scale $\Delta \rho_\chi$ is the difference of density across the inner core due to either thermal or compositional stratification.
The quantity $\chi$ is scaled by $\Delta \rho_\chi /\alpha \rho$. 
The characteristic
velocity scale is defined using the diffusion time scale rather than the inner core growth rate, to make it usable in both 
the growing and non-growing inner core cases. The quantity
$S(t)$ is made dimensionless using $r^2_\mathrm{ic} \alpha \rho /\kappa \Delta \rho_{\chi}$. 
Using the same symbols for the dimensionless quantities (including using now
$r$ for the dimensionless radius $\tilde{r}$ defined in the last subsection), we obtain
  \begin{equation}
  \mitbf{\nabla}\cdot \mitbf{u}=0,\label{eq:dimensionlessSetofEquations1}
\end{equation}
\begin{equation}
  \mitbf{0}=-\mitbf{\nabla}p'+Ra(t)\, \chi \,r\, \mitbf{e}_r+\nabla^2 \mitbf{u}+M(t)\mitbf{f}_L,\label{eq:dimensionlessSetofEquations2}
\end{equation}
\begin{multline}
  \xi(t) \derpar{\chi}{t}=-\left ( \mitbf{u}-Pe(t)\, r\, \mitbf{e}_r \right )\cdot \mitbf{\nabla}\chi +\nabla^2 \chi 
  \\+S(t)-\chi \xi \frac{\dot{\Delta \rho_\chi}}{\Delta \rho_\chi},
     \label{eq:dimensionlessSetofEquations3} 
\end{multline}
with four dimensionless parameters defined as
\begin{eqnarray}
    Ra(t)&=&\frac{\Delta \rho_\chi(t) g_{ic} r_{ic}^3(t)}{\eta \kappa},\label{eq:parameters_Ra}\\
    M(t)&=&\frac{B_0^2r^2_{ic}(t)}{{\mu_0 \eta \kappa}},\label{eq:parameters_M}\\
    \xi (t)&=&\frac{r_{ic}^2(t)}{\kappa \tau_{ic}},\label{eq:parameters_xi}\\
    Pe(t)&=&\frac{u_{ic}r_{ic}(t)}{\kappa}.\label{eq:parameters_Pe}
\end{eqnarray}
The last term in Eq.~(\ref{eq:dimensionlessSetofEquations3}) comes from the time
evolution of the  scale for $\chi$, $\Delta
  \rho_\chi /\alpha \rho$.

$\xi(t)$ and $Pe(t)$ characterize the growth of the inner
core. The P\'{e}clet number $Pe(t)$ compares the apparent advection from
 the moving boundary to
diffusion. A large P\'{e}clet number thus corresponds to a fast inner core growth  compared to
the diffusion rate. In the case
of a non-growing inner core, $Pe=0$, 
$\dot{S}(t)=0$ and
the relevant timescale is no longer $\tau_{ic}$ but the diffusion time scale, which gives 
$\xi=1$. This approach allows us to treat both non-growing and growing cases with the same
set of dimensionless parameters.   

$M(t)$ is an effective Hartmann number, which
quantifies the ratio of the Lorentz
force to the viscous force, using thermal diffusivity in the velocity scale. This effective
Hartmann number is related to the Hartmann number often used in magnetohydrodynamics \citep{Roberts2007-TG},
$Ha=Br/\mu_0\eta \lambda$, through
$M=Ha^2\, \lambda/\kappa $, where $\lambda$ is the magnetic diffusivity.

 $Ra(t)$ defined in equation \eqref{eq:parameters_Ra} is the Rayleigh number that
 characterizes the   stratification, and is
 negative since $\Delta \rho_{\chi}$ is negative for a stable stratification. 
The density stratification depends on the importance of diffusion, and on the time-dependence of the inner core radius.
Expressions for $\Delta \rho_{T}$ and $\Delta \rho_{c}$ will be given in section \ref{sec:simpl-growth-inner}.

To solve numerically the momentum equation \eqref{eq:dimensionlessSetofEquations2}, the
velocity field is decomposed into  poloidal and  toroidal components. The complete treatment of this
equation and the expression of the Lorentz 
force in term of poloidal and toroidal decomposition are described in appendix~\ref{sec:poloidal-form-set}.

\subsubsection{Simplified growth of the inner core}
\label{sec:simpl-growth-inner}
 
A realistic model for the inner core thermal evolution can be obtained by
calculating the time evolution of the source term $S_T(t)$ and the radius $r_{ic}(t)$ from
the core energy balance \citep{Labrosse:2003p150,Labrosse2015}, as done by \citet{Deguen:2011ga}.  
The result is sensitive to
the physical  properties of the core.
To focus on the effect of the Lorentz forces,  we choose a simpler  growth scenario and assume that 
the inner core radius increases as the square root of time \citep{Buffett:1992vk}. 
Using $r_{ic}(t)=r_{ic}(\tau_{ic}) (t/\tau_{ic})^{1/2}$ with
$r_{ic}(\tau_{ic})$ the present  radius of the inner core, the
growth rate is thus $u_{ic}(t)=r_{ic}(\tau_{ic})/2\sqrt{\tau_{ic} t}$.

This leads to the following expressions for the control parameters:
\begin{eqnarray}
Ra(t)&=&Ra_0 \, \frac{\Delta \rho_\chi (t)}{\Delta \rho_{\chi,0}}\, t^{3/2},\label{eq:parameters_functionoft_Ra_chi} \\
  M(t) & = & M_0\, t,\label{eq:parameters_functionoft_M}\\
  \xi(t)&=&2 \, Pe_0\, t,\label{eq:parameters_functionoft_xi}\\
  Pe(t)&=&Pe_0, \label{eq:parameters_functionoft_Pe}
\end{eqnarray}
where the subscript $0$
corresponds to 
values for the present inner core, and $t$ is dimensionless.

The P\'{e}clet number $Pe(t)$ is constant and equal to $Pe_0=r_\mathrm{ic}^{2}(\tau_{ic})/(2 \kappa \tau_{ic})$,
and the parameter $\xi$ is proportional to $Pe_0$. We are left with only
three independent dimensionless parameters: the Rayleigh number $Ra_0$
characterizes the density stratification, the effective Hartman number $M_0$ the strength of the
magnetic field, and the P\'eclet number $Pe_0$
the the relative importance of advection from the growth of the inner core and diffusion.

The value and time dependence of $\Delta\rho_\chi (t)$ depends on whether a stratification of thermal or compositional origin is considered:
\begin{itemize}
\item
In the thermal case, the source
term for thermal stratification  $S_T(t)$ defined in Eq.~(\ref{eq:St}) can also be written as 
\begin{equation}
\label{eq:S(t)}
 S_{T}(t) = \frac{\rho g' \gamma T}{K_S} \left[ \left( \frac{dT_\mathrm{s}}{dT_\mathrm{ad}}-1 \right)  r_\mathrm{ic}(t) u_\mathrm{ic}(t) - 3 \kappa_T \right], 
\end{equation}
where $dT_s/dT_{ad}$ is the  ratio of the Clapeyron slope to the adiabat
gradient, $g'=dg/dr=g_{ic}/r_{ic}$, $\gamma$ the Gruneisen parameter, and $K_S$  the  isentropic bulk modulus \citep{Deguen:2011ga}. 
With $r_{ic} \propto t^{1/2}$, the product 
$r_{ic}(t)u_{ic}(t)$ is constant, and so is $S_T$. 

Solving the energy conservation equation for the potential temperature ($\chi=\Theta$) assuming $\mitbf{u}=\mitbf{0}$, $r_\mathrm{ic}\propto t^{1/2}$, and $S_{T}$ constant gives
\begin{equation}
\Theta = \frac{S_{T} r_{ic}^2}{6\kappa_T (1+Pe_{T0}/3)} \left[1-\left(\frac{r}{r_{ic}(t)}\right)^2\right]	\label{eq:Theta_r}	
\end{equation}
in dimensional form (see appendix~\ref{sec:therm-strat} for the derivation). If $Pe_{0}\ll 1$, then the potential temperature difference $\Delta \Theta$  across the inner core is $S_{T} r_{ic}^2/6\kappa $, which corresponds to a balance between  effective heating ($S_{T}$) and diffusion.
In contrast, $\Delta \Theta$ tends toward $S_{T}  \tau_{ic}$ if diffusion is ineffective and $Pe_{0}\gg 1$.
From Eq. \eqref{eq:Theta_r}, we obtain 
\begin{equation}
\Delta \rho_{T} = \frac{\alpha_{T}\rho\, S_{T} r_{ic}^2}{6\kappa_T (1+Pe_{T0}/3)}
\end{equation}
and
\begin{equation}
Ra_{T} = \frac{\alpha_{T} \rho\, g_{ic} S_{T} r_{ic}^{5}}{6 \eta \kappa_T^{2}(1+Pe_{T0}/3)}.   
\end{equation}
With $g_{ic}\propto r_{ic}$ and $r_\mathrm{ic}\propto t^{1/2}$, this gives $Ra_{T}\propto r_{ic}^{6} \propto t^{3}$ and 
\begin{equation}
  Ra_T(t) =  Ra_{T0}\, t^3.\label{eq:parameters_functionoft_Ra}
\end{equation}

\item
We estimate the density stratification due to composition 
from the equation of solute conservation, assuming that the outer core is well-mixed and that the partition coefficient is constant.
The compositional P\'eclet number is large ($Pe_{C}\sim 10^{5}$
  with a diffusivity $\kappa_{C}\sim 10^{-10}$ m.s$^{-2}$) and solute diffusion in the inner core is therefore neglected.

The composition of the solid crystallized at time $t$ at the ICB is estimated as
  \begin{equation}
  \label{eq:concentration_solid_ICB}
  c^s_{icb}(t)=kc_0^l \left ( 1-\left ( \frac{r_{ic}(t)}{r_c}\right )^3\right
  )^{k-1}
\end{equation}
(see Appendix~\ref{sec:comp-strat}), from which the density difference across the inner core is
  \begin{eqnarray}
\Delta \rho_{C}(t) &=& \alpha_{C} \rho \left[ c^s_{icb}(t) - c^s_{icb}(t=0) \right],\\
&=& \alpha_{C} \rho kc_0^l \left[\left ( 1-\left ( \frac{r_{ic}(t)}{r_c}\right )^3\right  )^{k-1} -1 \right].
\end{eqnarray}
We take advantage of the smallness of $(r_{ic}(t)/r_{c})^{3}<4.3\, 10^{-2}$ to approximate $\Delta \rho_{C}$ as
\begin{equation}
\Delta \rho_{C}(t) \simeq \alpha_{C} \rho k(1-k)c_0^l \left(\frac{r_{ic}(t)}{r_c}\right )^3,
\end{equation}
which gives
\begin{equation}
Ra_{C}=\frac{\Delta \rho_C(t) g_{ic} r_{ic}^3(t)}{\eta \kappa_{C}}=\frac{\alpha_{C} \rho k(1-k)c_0^l g_{ic} r_{ic}^6(t)}{\eta \kappa_{C} r_c^{3}}.
\end{equation}
With $g_{ic}\propto r_{ic}$ and $r_\mathrm{ic}\propto t^{1/2}$, this gives $Ra_{C}\propto r_{ic}^{7} \propto t^{7/2}$ and
\begin{equation}
Ra_C(t)=Ra_{C0}\,t^{7/2}. 
\end{equation}
\end{itemize}

  \begin{table*}
    \caption{Typical values for the parameters used in the text, and typical range of
      values when useful.}
      \begin{minipage}{100mm}
    \begin{tabular}{@{}|l|c|c|c|@{}}
        Parameter& Symbol & Typical value & Typical range\\
        \hline
        Magnetic field&$B_0$& $3\times 10^{-3}$\,T& $10^{-1}-10^{-3}$\,T\\
        Thermal diffusivity$^c $ &$\kappa_T$&$1.7\times 10^{-5}$\,m$^2$.s&$0.33-2.7\times 10^{-5}$\,m$^2$.s\\
        Chemical diffusivity$^a$  &$\kappa_{\chi }$& $10^{-10}$\,m$^2$.s&$10^{-10}-10^{-12}$\,m$^2$.s\\
        Viscosity&$\eta$&$10^{16}$ Pa.s&$10^{12}-10^{21}$ Pa.s\\
        Age of IC&$\tau_{ic}$&$0.5$ Gyrs&$0.2-1.5$ Gyrs\\
        Density stratification (thermal case)$^d$&$\Delta \rho_T$&$6$\,kg.m$^{-3}$&$0.5-25$\,kg.m$^{-3}$\\
        Density stratification (compositional case)$^e$ &$\Delta \rho_C$&$5$\,kg.m$^{-3}$&$1-10$\,kg.m$^{-3}$\\
        Phase change timescale&$\tau_{\phi}$&$10^3\,$\,yrs&$10^2$-$10^4$\,yrs\\
        \hline
        Inner core radius$^b$&$r_{ic}(\tau_{ic})$&$1221\,$\,km&\\
        Acceleration of gravity ($r=r_{ic}$ &$g_{ic}$&$4.4$\,m.s$^{-2}$&\\
        Density of the solid phase$^b$ &$\rho$&$12800$\,kg.m$^{-3}$&\\
        Density difference at the ICB&$\delta \rho_{ic}$&$600$\,kg.m$^{-3}$&\\
        Thermal expansivity &$\alpha$&$10^{-5}$\,K$^{-1}$&\\
        Permeability &$\mu_0$&$4 \pi \times 10^{-7}$\,H.m$^{-1}$&\\
        \hline
      \end{tabular}
      \label{tab:values}
    \end{minipage}

\medskip
\begin{flushleft}

  $^a$ from \citet{Gubbins:2013ip}

  $^b$ from PREM \citet{Dziewonski:1981p60}

  $^c$ obtained using $k=163$\,W.m$^{-1}$.K$^{-1}$, $c_p=750$\,J.K$^{-1}$.kg$^{-1}$
  \citep{Pozzo:2012ty,Gomi:2013jp}

  $^d$ assuming $S=10-1000$\,K.Gyrs$^{-1}$ \citep{Deguen:2011ga}

  $^e$from \citet{Deguen:2011ga}
\end{flushleft}

  \end{table*}

  \begin{table*}
 \caption{Typical values of the dimensionless parameters discussed
      in the text for the present inner core, using typical values from table
      \ref{tab:values}. }
\label{tab:dimensionlessparameters}
\def\arraystretch{2.2}
    \begin{tabular}{@{}lcll@{}}
      \hline
      Dimensionless parameter &Symbol&Thermal&Compositional\\
      \hline
      Rayleigh number & $Ra$&$\dfrac{10^{16}\,\textrm{Pa.s}}{\eta}\quad \times(-2.8 \times
      10^8)$&$\dfrac{10^{16}\,\textrm{Pa.s}}{\eta}\quad \times ( {-8 \times 10^{12}})$\\
      effective Hartmann number &$M$&$\left ( \dfrac{B_0}{3\times 10^{-3}\,\textrm{T}}\right )^2
      \dfrac{10^{16}\,\textrm{Pa.s}}{\eta} \quad \times  {63}$&$\left ( \dfrac{B_0}{3\times
          10^{-3}\,\textrm{T}}\right )^2 \dfrac{10^{16}\,\textrm{Pa.s}}{\eta} \quad \times  {1.07\times 10^7}$\\
      P\'eclet number &$Pe$&$ {2.8}$&$ {4.7 \times 10^5}$\\
      Phase change number &$\mathcal{P}$&$\dfrac{10^{16}\,\textrm{Pa.s}}{\eta}\quad
      \times  {10^4}$&$\dfrac{10^{16}\,\textrm{Pa.s}}{\eta}\quad \times {10^4}$\\
      \hline
    \end{tabular}
\medskip

See the definitions of the dimensionless parameters in the
      text.  
  \end{table*}

\subsection{Boundary conditions}
\label{sec:boundary-conditions}

The Earth's inner core boundary is defined by the coexistence of solid and liquid iron, at the temperature of the liquidus for the given pressure and composition.   
By construction, the potential temperature $\Theta$ and  concentration $C$ are both 0 at the ICB : $\Theta(r_{ic}(t)) =  C(r_{ic}(t))=0$.
The mechanical boundary conditions are tangential stress-free conditions and continuity of the normal stress at the inner core boundary. 

When allowing for phase change at the ICB,
the  condition of continuity of the normal stress gives
\begin{equation}
  \label{eq:continuitynormalstress}
  -\mathcal{P}(t)(u_r-u_{ic})-2 \derpar{u_r}{r}+p'=0
\end{equation}
in dimensionless form. 
The parameter $\mathcal{P}(t)$  was introduced by \citet{Deguen:2013bj} to
characterize the phase change, and is the ratio of the phase change timescale
$\tau_{\phi}$ to the viscous relaxation timescale $\tau_{\eta}=\eta/(\delta \rho\, g_{icb}r_{ic})$,
\begin{equation}
  \label{eq:Pt}
  \mathcal{P}(t)=\frac{\tau_{\phi}\delta \rho\, g_{icb}r_{ic}}{\eta}, 
\end{equation}
where $\delta \rho$ is the density difference between liquid and solid iron at the inner core
boundary. $\tau_\phi$ has been estimated to be $\sim 10^3$\,years \citep{Alboussiere:2010p24,Deguen:2013bj}.
The limit $\mathcal{P}\to 0$ corresponds to  perfectly permeable boundary conditions where the
phase change occurs instantaneously, and $\mathcal{P}\to \infty $ corresponds to  perfectly impermeable boundary
conditions with no phase change allowed at the boundary. 

With $r_{ic}(t)\propto t^{1/2}$ and $\tau_\phi$ constant, 
the parameter $\mathcal{P}(t)$ is expressed using the current value 
$\mathcal{P}_0=\mathcal{P}(t=\tau_{ic})$ as
\begin{equation}
  \label{eq:P(t)}
  \mathcal{P}(t)=\mathcal{P}_0\, t.
\end{equation}

\subsection{Numerical modelling}
\label{sec:numerical-modelling}

The code is an extension of the one used in \citet{Deguen:2013bj}, adding the effect of
the magnetic forcing as in Eq.~(\ref{eq:momentumconservation}). The system of
equations derived in appendix~\ref{sec:poloidal-form-set} in term of poloidal/toroidal
decomposition is solved in axisymmetric geometry, 
using a
spherical  harmonic expansion for
the horizontal dependence and a finite difference scheme in the radial direction.  The
nonlinear part of the advection term in the temperature (or composition) equation is
evaluated in the
physical space at each time step.  A semi-implicit Crank-Nicholson  scheme is implemented
for the time evolution of the linear terms and an Adams-Bashforth procedure   is used for
the nonlinear  advection term in the heat equation.

The boundary conditions 
are the same as in \citet{Deguen:2013bj}, but for most of the runs we use $\mathcal{P}=10^6$, which correspond to
impermeable boundary conditions as discussed in section~\ref{sec:boundary-conditions}. 

When keeping the inner core radius constant, the code is run until 
a steady state is reached. Otherwise, the code is run from $t=0.01$ to $t=1$.

\section{Flow description}
\label{sec:flow-description}

\subsection{Neutral stratification}
\label{sec:neutr-strat}

In this subsection, we
investigate the effect of the boundary conditions on the geometry and strength of the flow by solving analytically the set of equations in the case of neutral stratification. The analytical
solution for neutral stratification has also been used to benchmark the code for $Ra=0$.

\begin{figure}
  \centering

  \includegraphics{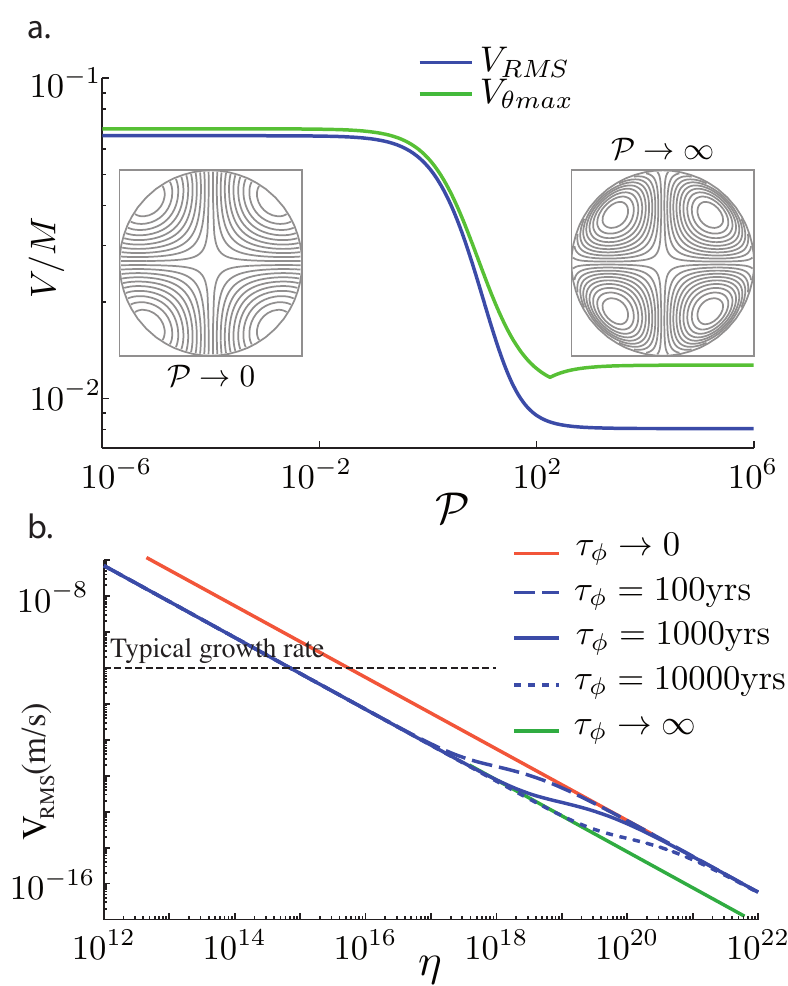}
 
  \caption{Analytic solution for $Ra=0$. (a) Evolution of the dimensionless velocity as a
    function of the phase change number $\mathcal{P}$, with streamlines for $\mathcal{P}\to
    0$ (left) and $\mathcal{P}\to \infty$ (right). The RMS velocity and the maximum of the
    horizontal velocity are plotted. (b) Evolution of the RMS velocity as a
  function of $\eta$, with velocity in m.s$^{-1}$. Except for the viscosity and the phase
  change time scale $\tau_\phi$,  the parameters used for definition of $\mathcal{P}$ and $M$
are given in Table~\ref{tab:values}. The kink in the curves corresponds to the change in
regime between large $\mathcal{P}$ (low viscosity) and low $\mathcal{P}$ (large
viscosity), and the corresponding viscosity value is a function of the phase change timescale $\tau_\phi$.}
  \label{fig:analyticsolution}
\end{figure}

In the case of neutral stratification, with $Ra=0$,  the equations for the
temperature or composition
 perturbation~(\ref{eq:dimensionlessSetofEquations3}) and momentum conservation~(\ref{eq:vorticity_equa})  are no longer coupled. The diffusivity is no longer relevant and the problem does
not depend on the P\'{e}clet number. 
Eq.~(\ref{eq:vorticity_equa}) is  solved in appendix~\ref{sec:analyt-solut-ra=0} using the boundary conditions presented in the 
previous section. The flow velocity is found to be proportional to the effective Hartman number $M$ times a sigmoid function of $\mathcal{P}$.
The  solution is shown on Fig.~\ref{fig:analyticsolution},
with dimensionless  maximum horizontal velocity and root mean square velocity as
functions of the phase change number $\mathcal{P}$, as well as streamlines corresponding to
the two extreme cases, $\mathcal{P}=0$ (fully permeable boundary conditions) and
$\mathcal{P}\to \infty$ (impermeable boundary conditions).

In the limit $\mathcal{P}\ll  1$, corresponding to permeable boundary conditions, the streamlines of the flow cross the ICB, which indicates significant melting and freezing  at the ICB. 
In contrast, the streamlines in the limit  $\mathcal{P}\gg  1$ are closed lines which do not cross the ICB, which indicates negligible melting or freezing at the ICB.
The velocity is
proportional to  the effective Hartmann number $M$ whereas the $\mathcal{P}$
dependence is more complex. The velocities reaches two asymptotic values for low and large
$\mathcal{P}$ values, separated by a sharp kink.  The discontinuity in the derivative of the maximum horizontal
velocity slightly above $\mathcal{P}\sim  10^2$ corresponds to a change of the spatial position of the
maximum, when the streamlines become closed and the maximal horizontal velocity is
obtained at the top of the cell and no longer at its bottom. 
The  change of behavior of the boundary from permeable to impermeable induces a significant decrease of the strength of the flow, since the velocity magnitude in the $\mathcal{P}\gg 1$ regime is one order of magnitude smaller than when permeable boundary conditions ($\mathcal{P}\ll 1$) are assumed.

Fig.~\ref{fig:analyticsolution}.b shows the maximum of the velocity, now given in m.s$^{-1}$,  as a function of the
viscosity, using typical values of the parameters given in Table~\ref{tab:values} and five different values for the  phase change timescale $\tau_\phi$,
from zero to infinite.  In term of dimensionless parameters, a high viscosity
corresponds to small Hartmann number $M$ and phase change number $\mathcal{P}$. Changing
the timescale $\tau_\phi$ translates the
position of the transition between the two regimes, the viscosity value corresponding to the transition being proportional to $\tau_{\phi}$, but does not change the general trend of
the curve, which is a linear decrease of the velocity magnitude in  log-log space, except for the kink between
the two regimes. The linear decrease is due to the viscosity dependence of the
  Hartmann number $M\propto \eta^{-1}$. 
For typical values of the  phase change timescale between 100 years and $10\,000$ years,
the kink
between the two regimes occurs at a viscosity in the range $10^{17}-10^{21}$ Pa.s.

In what follows, we will focus on the conditions which are the most favorable to deformation due to the poloidal component of the Lorentz force, 
 and therefore focus on the case of  low viscosity and large $\mathcal{P}$. 
    The ICB would act as a permeable boundary only if $\mathcal{P}\lesssim 10^{2}$ (see Fig. \ref{fig:analyticsolution}), corresponding to $\eta \gtrsim 10^{17}$ Pa.s.
  Under these conditions, the typical flow velocity would be  $\lesssim 10^{-12}$ m.s$^{-1}$, \textit{i.e.} two orders of magnitude or more smaller than the inner core growth rate, and would be unlikely to result in significant texturing. 
  For this reason, we will let aside the high viscosity/low $\mathcal{P}$ regimes to focus on low viscosity/high $\mathcal{P}$ cases, for which the ICB acts as an impermeable boundary.   
This gives  boundary conditions very different from previous studies, where perfectly permeable boundary conditions were assumed \citep{Karato:1999tt,Buffett:2000tb}. 
In particular, this implies that the flow velocity estimated by \citep{Karato:1999tt} was overestimated by one order of magnitude.

According to Eq.~(\ref{eq:P(t)}), the parameter $\mathcal{P}$ varies linearly with time,
which means that $\mathcal{P}$ must have been small early in inner core's history.
However, this is true for a very short time, when the inner core radius was very small, of
the order $r_{ic}(\tau_{ic})/\mathcal{P}_{0}^{1/2}$, and this episode is unlikely to have
observable consequences in the present structure of the inner core.

\subsection{Zero growth rate}
\label{sec:zero-growth-rate}

\begin{figure}
  \centering
  \includegraphics{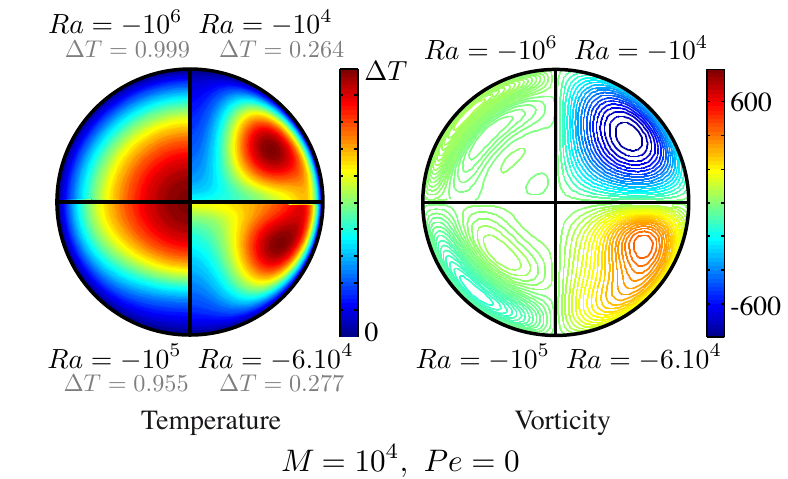}
  \caption{Snapshots of meridional cross section of the temperature and the vorticity fields 
    for  $M=10^4$ and a constant inner core radius,  for four different values of the Rayleigh number (from top right,
    going clockwise: $Ra=-10^4$,
    $-6\times 10^{4}$, $-10^5$, $-10^6$).  When the stratification is large enough
    ($Ra=-10^{6}$), the flow is confined at the top of the inner core and the temperature
    field has a spherical symmetry. When the stratification is weak ($Ra=-10^4$), the flow
  is similar to the one in Fig.~\ref{fig:analyticsolution} for $Ra=0$ and the
  temperature is almost uniform. The vorticity is scaled by $\kappa_T /r_{ic}^2$ and  the
  temperature by $Sr^2_{ic}/6\kappa_T$. For $u_\mathrm{ic}=0$,  $Sr^2_{ic}/6\kappa_T$ reduces to
  $T_s(0)-T_s(r_\mathrm{ic})$.
}
  \label{fig:phenom_nogrowth}
\end{figure}

We first investigate the effect of the Lorentz force without taking into account the
secular growth of the inner core ($Pe=0$). 
Fig.~\ref{fig:phenom_nogrowth} shows the vorticity and temperature fields obtained for different values
of the Rayleigh number, at a given effective Hartmann number $M=10^4$, for a thermally stratified
inner core. 

When
the Rayleigh number is small, the vorticity field is organized in two symmetric tores wrapped
around the N-S axis.The stratification is too weak to alter the flow induced by
the Lorentz force  and the temperature field is advected and
mixed by the flow. 
The velocity field is equal to the one calculated analytically for $Ra=0$ (see
subsection~\ref{sec:neutr-strat} and
appendix~\ref{sec:analyt-solut-ra=0}). 

However, when the Rayleigh number is larger, the flow is altered by the stratification and is confined in an uppermost layer, as found by \citet{Buffett:2000tb}. 
The velocity is smaller than in the case of neutral stratification.  The temperature field
is strongly stratified and the perturbations due to radial advection are small. 
The flow obtained here is similar to the flow induced by differential inner core growth
with a stable stratification \citep{Deguen:2011ck}, with a notable
difference: we impose  a large $\mathcal{P}$ implying a near zero
radial flow $v_r$ across the ICB, whereas \citet{Deguen:2011ck}
impose a given $v_r$ as the driving force. 
The confinement of the flow in a thin layer is likely to
concentrate the deformation and thus we may expect higher strain rates for a
highly stratified inner core, but a different spatial distribution of the deformation.

\begin{figure}
  \centering
\includegraphics{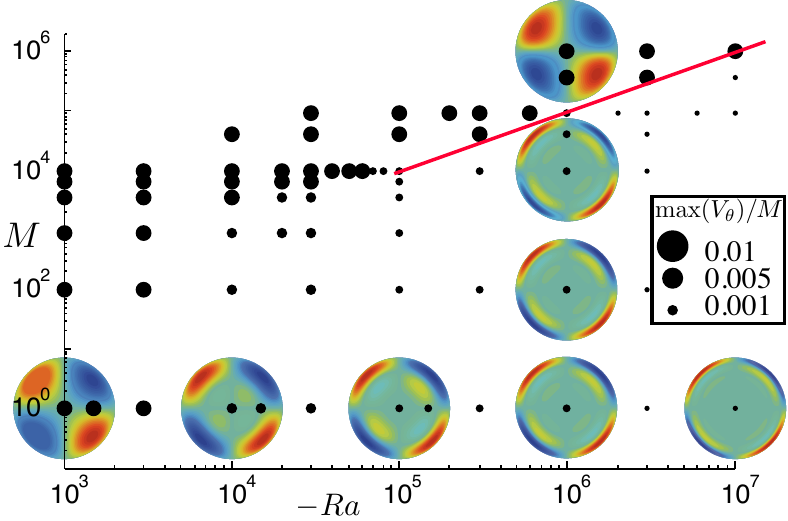}
  \caption{Maximum velocity (normalized by the Hartmann number $M$) in the upper vorticity layer, for a zero growth rate. The
    velocity is scaled by the diffusion velocity $\kappa/r_\mathrm{ic}$. The maximum
    size of the dots corresponds to the value for $Ra=0$ computed analytically.  For some values of $(M, -Ra)$, the vorticity field is 
    plotted in the meridional cross section. The red line with a slope of $1$ shows the
    limit between the two regimes. }
  \label{fig:M_Ra}
\end{figure}

To explore the parameter space in terms of Rayleigh and effective Hartmann numbers, we computed runs  with Rayleigh numbers from $-10^3$ to $-10^7$ and effective Hartman number from
$10^0$ to $10^6$. The maximum velocity (normalized by $M$) is plotted in Fig.~\ref{fig:M_Ra} as a proxy to
determine the regime. The largest velocity coincides with the flow velocity obtained for neutral
stratification. The vorticity field corresponding to some of the points in the regime diagram
are also shown in Fig.~\ref{fig:M_Ra}. 

The systematic exploration of the parameter space reveals two
  different dynamical regimes, which domains of existence in a ($-Ra$,$M$) space are shown in Fig. \ref{fig:M_Ra}.
In the
upper left part of the diagram (large effective Hartmann number, low Rayleigh number),  the flow is very similar (qualitatively and quantitatively) to the analytical solution for a neutral
stratification, and deformation extends deep in the inner core.
This regime is characterized by a negligible effect of the buoyancy
forces, and will therefore be referred to as the \textit{weakly
  stratified} regime. 
In the lower right part, the flow is confined in a shallow layer which
thickness depends on the Rayleigh number only (not on $M$) and in which the velocity is smaller than for neutral stratification. 
This regime will be referred to as the \textit{strongly stratified} regime.

\subsection{Growing inner core}
\label{sec:growing-inner-core}

To investigate the effect of  inner core growth, we compute several runs with a
given set of parameters $(Ra_0,M_0,Pe_0)$, with the time $t$
between $t=0.01$ and $t=1$. 
Unlike in subsections~\ref{sec:neutr-strat} and \ref{sec:zero-growth-rate}, the dimensionless numbers evolve with 
time, as described by Eqs~\eqref{eq:parameters_functionoft_Ra_chi} to \eqref{eq:parameters_functionoft_Pe}.

\begin{figure*}
  \centering
  \includegraphics{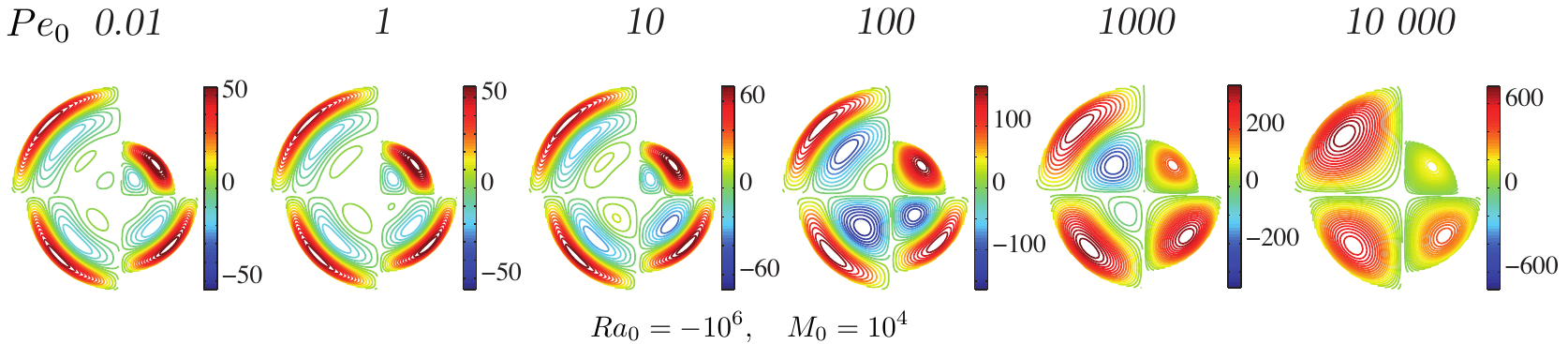}
\caption{Snapshots of the vorticity field for simulations with dimensionless parameters $M_{0}=10^4$, $Ra_{0}=-10^6$,   
and $Pe_{0}=0.01$, 1, 10, $10^{2}$, $10^{3}$ and $10^{4}$ (from left to right), with $r_{ic} \propto t^{1/2}$. Each panel corresponds to one simulation, with four time steps
  represented : $t=0.25$, $0.50$, $0.75$ and $1$ dimensionless time, from top-right and
  going clockwise.  See Fig. \ref{fig:vonMises_vorticity} for strain rates of corresponding
runs.}
  \label{fig:timeseries_vorticity}
\end{figure*}

Fig.~\ref{fig:timeseries_vorticity} shows the evolution of the vorticity field in six
simulations, for a thermal stratification, with the same Rayleigh and effective Hartman numbers, $Ra_{T0}=-10^6$, $M_{T0}=10^4$, but different values of the P\'{e}clet number, 
which corresponds to increasing diffusivity from left to right. 
For each run,
snapshots of the vorticity field corresponding to four time steps are shown, from top-right and going clockwise.

Fig.~\ref{fig:timeseries_vorticity} shows that the thickness of the upper layer increases with increasing P\'eclet number.
The transition between the two regimes
of strong and weak stratification is shifted toward larger Rayleigh
numbers when the P\'eclet number is increased. 
At low or moderate P\'eclet numbers ($Pe_0\leq 10^{2}$
in the cases presented here), the magnitude of vorticity is almost constant time, implying that the
deformation rate in the uppermost layer is also constant. 

\section{Scaling laws}
\label{sec:scal-laws-numer}

In this section, we determine scaling laws  for the thickness of the shallow shear layer and the
maximum velocity in the layer in the strongly stratified regime from the set of equations developed in section~\ref{sec:governing-equations}. 
We will first discuss the transition between the strongly stratified and weakly stratified regimes
discussed in Fig.~\ref{fig:M_Ra}.
We will then focus on the strongly stratified regime
and estimate the deformation in the uppermost layer. Thermal
and compositional stratification are discussed separately.   
The flow in the weakly stratified regime is given by the analytical model discussed in section~\ref{sec:neutr-strat} and appendix~\ref{sec:analyt-solut-ra=0} for neutral stratification.

\subsection{Balance between magnetic forcing and stratification}
\label{sec:boundary-between-two}

We start here by discussing the transition between the strongly-stratified and 
weakly-stratified regimes.
We base our analysis on the vorticity equation obtained by taking the curl of the momentum conservation equation (Eq.~(\ref{eq:dimensionlessSetofEquations2})),
\begin{equation}
    \label{eq:vorticity_equa}
    \mitbf{0}=-Ra(t) \derpar{\chi}{\theta}\mitbf{e}_{\phi } +M(t) \mitbf{\nabla}\times \mitbf{f}_L+\nabla ^2
    \mitbf{\omega},  
  \end{equation}
where $\mitbf{\omega}=\mitbf{\nabla}\times \mitbf{u}$ is the vorticity.  
The quantity   $\chi $ (denoting either potential temperature or composition) is split into two parts, $\chi=\bar{\chi}(r,t)+\chi '(r,\theta,t)$, where
$\bar{\chi}$ is the reference radial profile corresponding to $\mitbf{u}=\mitbf{0}$.
The vorticity equation then writes
 \begin{equation}
 \mitbf{0}=Ra \derpar{\chi '}{\theta}\mitbf{e}_{\phi} +M\, \mitbf{\nabla}\times \mitbf{f}_L+\nabla ^2 \mitbf{\omega}.
   \label{eq:systemeq11}
 \end{equation}
In the vorticity equation, the three terms must balance if the effect of stratification is important. 
Starting from a state with no perturbations,  $\chi'=0$, the flow velocity is initially set by a balance between the Lorentz force and
the viscosity force. Isosurfaces of $\chi$ are deformed by the resulting flow, and the buoyancy force
increases,  eventually balancing the magnetic force if the stratification is strong enough. 
In this case, further radial motion is prevented and the flow tends to be localized in a  layer below the ICB, as found in our numerical simulations.  
Denoting by $\delta$ the thickness of the
shear layer and  $u$ and $w$   the  horizontal and vertical velocity, respectively, the vorticity is $\omega \sim  u/ \delta$. We therefore have, from Eq.~\eqref{eq:systemeq11},
\begin{equation}
  \label{eq:vortequ}
  (-Ra) \chi '\sim M \sim \frac{u}{\delta^3}. 
\end{equation}
The  perturbation $\chi'$ thus scales  as
\begin{equation}
  \label{eq:temperature_perturbations}
  \chi '\sim \frac{M}{-Ra}
\end{equation}
if the stratification is strong enough for the induced buoyancy forces to balance the Lorentz force.

The effect of the stratification is negligible if the buoyancy forces,
which are $\sim - Ra \chi '$, cannot
balance the Lorentz force, which is $\sim M$. 
Since $\chi '$ is necessarily smaller than $|\bar{\chi}(r_{ic}) - \bar{\chi}(0)|$,
which by construction is equal to 1,  the effect of the stratification will therefore be
negligible if $M \gg    -Ra$.
This is consistent with the boundary between the two regimes found from our numerical
calculations, as shown in  Fig.~\ref{fig:M_Ra}, as well as with the results of
\citet{Buffett:2000tb} who found that the Lorentz force can displace isodensity surfaces
by  $\sim r_\mathrm{ic} M/(-Ra)$.
This estimate is valid for both a growing or non-growing inner core.

\subsection{Scaling laws in the strongly stratified regime} 
\label{sec:scal-laws-thickn}

The reference profile $\bar \chi$ is solution of 
\begin{equation}
  \label{eq:ThetaBar}
   \xi  \derpar{\bar{\chi}}{t}= \nabla^2 \bar{\chi} +Pe\, \mitbf{r}\cdot  \mitbf{\nabla} 
   \bar{\chi}+S(t)-\xi\frac{\dot{\Delta \rho_\chi }}{\Delta  \rho_\chi}\bar \chi.
 \end{equation}
Subtracting Eqs~(\ref{eq:ThetaBar}) to (\ref{eq:dimensionlessSetofEquations3}), and
 assuming that $\chi '\ll \bar \chi$, we obtain 
\begin{equation}
  \label{eq:scalingheat}
  \underbrace{\xi \derpar{\chi'}{t}}_{\sim \xi \chi '}= \underbrace{\nabla^2 \chi
    '}_{\sim \chi '/\delta ^2} -\underbrace{u \derpar{\chi'}{\theta}}_{\sim u\chi '}-\underbrace{w \derpar{\chi'}{r}}_{w \chi'/\delta '}-\underbrace{w
  \derpar{\bar{\chi}}{r}}_{\sim w\bar{\chi} \sim w}+\underbrace{Pe\, \mitbf{r}\cdot \mitbf{\nabla }\chi'}_{\sim Pe \chi '/\delta
} -\underbrace{\xi \frac{\dot{\Delta \rho}}{\Delta \rho} \chi '}_{\sim Pe \chi '}.
\end{equation}

Three of these terms depend on the growth rate: $\xi \partial \chi '/ \partial t
$,  $Pe\, \mitbf{r}\cdot \mitbf{\nabla}\chi'$, and $\xi  \, \dot{\Delta \rho }/\Delta \rho \chi '$. 
With our assumption of  $r_{ic}\sim
t^{1/2}$, we have $\xi=2 Pe \, t$ and thus $\xi \lesssim Pe$. Thus, the largest term among the growth rate-dependent terms is $Pe\, \mitbf{r}\cdot
\mitbf{\nabla} \chi ' \sim Pe\, \chi '/ \delta$.

Comparing the effect of the diffusion term, which is $\sim \chi '/\delta^2$, with the inner core
growth term, which is $\sim Pe \chi'/\delta  $, we find that the effect of the inner core growth is
negligible if 
\begin{equation}
  \label{eq:growtnegligible}
  Pe \ll \frac{1}{\delta}.
\end{equation}
This suggest the existence of two different regimes depending on whether $Pe$ is small or large.
We develop below scaling laws for these two cases. 

\subsubsection{Small $Pe$ limit }		
\label{sec:small-pe-limit}

Neglecting the growth terms, we have
\begin{equation}
  \label{eq:withoutgrowth}
  0=\underbrace{\nabla ^2 \chi '}_{\sim \chi '/\delta ^2}- \underbrace{u \derpar{\chi '}{\theta}}_{\sim u \chi '}-
  \underbrace{w \derpar{\chi '}{r}}_{w \chi '/\delta} -
  \underbrace{w \derpar{\bar{\chi}}{r}}_{\sim w \bar{\chi}}.
\end{equation}
The conservation of mass implies that $ u\sim  w / \delta$, and with  $\chi '\sim {M}/{Ra}$, we obtain 
\begin{equation}
  \label{eq:withoutgrowth2}
  0=\underbrace{\nabla ^2 \chi '}_{\sim  M /Ra \delta ^2}- \underbrace{u \derpar{\chi
      '}{\theta}}_{ \sim u M/Ra}-\underbrace{w \derpar{\chi '}{r}}_{uM/Ra}-
  \underbrace{w \derpar{\bar{\chi}}{r}}_{\sim u \delta }.
\end{equation}
We now assume that the  advection of the perturbation $\chi'$ is small compared to the vertical
advection of the reference state, which requires that $\delta \gg M/(-Ra)$. Balancing the advection and diffusion
terms, we obtain 
\begin{equation}
  \label{eq:hop}
  \frac{M}{(-Ra) \delta ^2}\sim u\delta. 
\end{equation}
Combining this expression with the relation $u/\delta ^3 \sim M$  obtained from the vorticity
equation (Eq.~\eqref{eq:vortequ}), we have
\begin{eqnarray}
  \delta \sim & (-Ra)^{-1/6},  \label{eq:SmallPelimit_delta}\\
u\sim & M \, (-Ra)^{-1/2}.   \label{eq:SmallPelimit_u}
\end{eqnarray}

\subsubsection{Large $Pe$ limit}
\label{sec:large-pe-limit}

In this limit, the diffusion time is larger than the age of the inner core, which allows us to neglect the
diffusion term. Keeping only  the largest growth rate-dependent term, and using  Eq.~\eqref{eq:vortequ},  we have
\begin{equation}
  \label{eq:largPe}
 0 =  -\underbrace{u \derpar{\chi'}{\theta}}_{~ u\chi '}-\underbrace{w \derpar{\chi'}{r}}_{ u\chi'}-\underbrace{w
  \derpar{\bar{\chi}}{r}}_{~ u\delta }+\underbrace{Pe\, \mitbf{r}\cdot \mitbf{\nabla
  }\chi'}_{~ Pe M /\delta Ra}.
\end{equation}
Assuming again that the advection of the perturbation is small compared to the vertical
advection of the reference state, the main balance is between the second and third terms, which gives
\begin{equation}
  \label{eq:largePebalance}
  u \, \delta \sim \frac{Pe \, M }{\delta \, (-Ra)}.
\end{equation}
Combined with  $u\delta\sim M \delta ^4$ from Eq.~\eqref{eq:vortequ}, 
we find that 
the thickness and maximum velocity of the upper layer are 
\begin{eqnarray}
  \delta &\sim & \left ( \frac{Pe}{-Ra} \right )^{1/5},  \label{eq:thickness_Pelarge_}\\
  u&\sim& M \left ( \frac{Pe}{-Ra} \right )^{3/5}. \label{eq:vorticity_Pelarge}
\end{eqnarray}

Two conditions have to be fulfilled for these scaling laws to be valid.  First, we must have $Pe \ll 1/\delta $, which is $Pe\gg (-Ra)^{1/6}$ using Eq.~\eqref{eq:thickness_Pelarge_}. Also,
we have assumed that the upper layer is thin ($\delta \ll  1$) and that		
the horizontal advection is small compared to the vertical one.

\subsubsection{Time-dependence}

Our derivation does not make any assumption on the form of the inner core growth $r_{ic}(t)$, and the scaling law validity should not be restricted to the $r_{ic}(t)\propto t^{1/2}$ case assumed in the numerical simulations.

These scaling laws are valid at all time during the growth of the inner core, provided that $Pe \ll 1/\delta$ and $\delta \ll 1$ and that the time dependence of the control parameters is properly taken into account.
For example, under the assumption of $r_{ic}\propto t^{1/2}$ and with $M$, $Ra$ and $Pe$ given by Eqs.~(\ref{eq:parameters_functionoft_M}), (\ref{eq:parameters_functionoft_Ra})
and (\ref{eq:parameters_functionoft_Pe}), we obtain 
\begin{equation}
  \label{eq:thicknessPeSmall_2}
  \delta \sim \left (-Ra_{T0}\right )^{-1/6} t^{-1/2},\quad u \sim  M_{T0}(- Ra_{T0})^{-1/2}  t^{-1/2}
\end{equation}
for the thermal case (small $Pe_{T}$), and
\begin{equation}
  \label{eq:thicknessPelarge_2}
  \delta \sim \left (\frac{Pe_{C0}}{(-Ra_{C0})}\right )^{1/5}  t^{-7/10},\quad  u \sim M_{C0}\left (\frac{Pe_{C0}}{(-Ra_{C0})}\right )^{3/5}  t^{-11/10}
\end{equation}
for the compositional case (large $Pe_{C}$).

\subsection{Comparison with numerical results}
\label{sec:check-numer-results}

\begin{figure}

  \centering
  \includegraphics{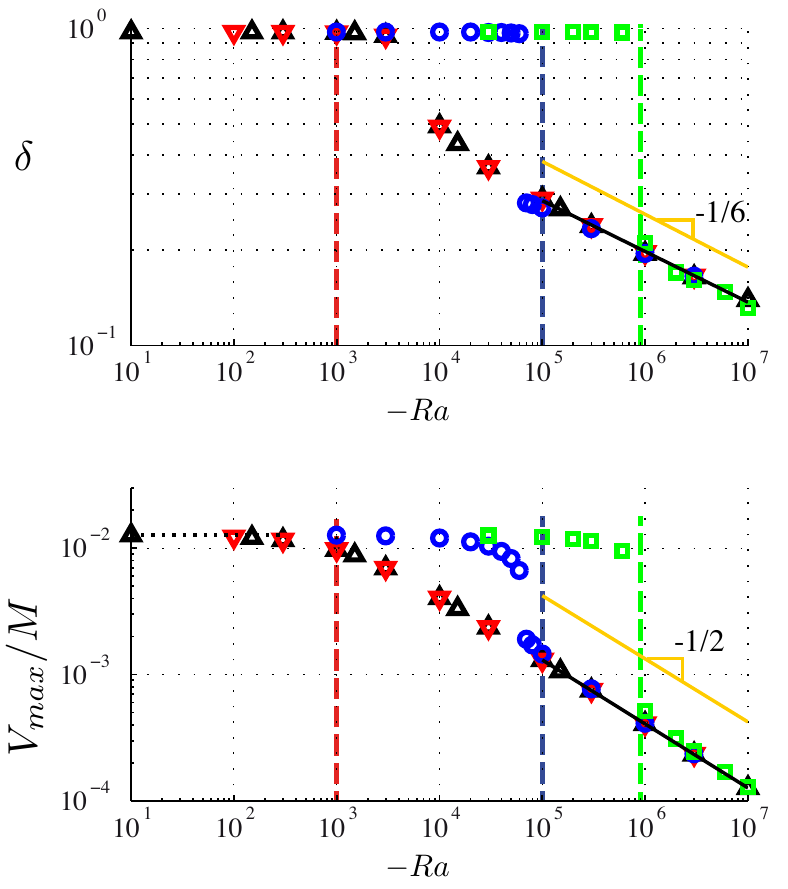}
  \caption{Results from simulations with a constant inner core radius. Evolution of the thickness of the uppermost vorticity
    layer (a) and  maximum horizontal velocity (b) with the absolute value of the
    Rayleigh number $-Ra$. Colors correspond to different values of the effective Hartmann number $M$, and dashed lines to
    the corresponding $-Ra=10 \,M$ line.  The velocity has been scaled with the effective Hartman number and
  the extreme value for low $-Ra$ corresponds to the analytical model with no stratification (black
  horizontal dotted line). The solid black lines are the best fit for $\delta<0.3$, 
  $\delta=1.81(-Ra)^{-0.16\pm0.013}$ and $V_\mathrm{max}=0.44M(-Ra)^{0.506\pm0.002}$. The orange
  lines show the slopes predicted in subsection \ref{sec:small-pe-limit}. }
  \label{fig:Statique}
\end{figure}

Fig.~\ref{fig:Statique} shows  the thickness of the uppermost vorticity layer and the maximum horizontal velocity obtained in numerical simulations with a constant inner core radius, which corresponds to the small P\'{e}clet number limit.  
When $-Ra/M\ll 1$, the flow has the   geometry and amplitude  predicted by our analytical model for
$Ra=0$. 
When $-Ra/M\gtrsim  10$ 
and the thickness of the upper layer is smaller
than $\simeq 0.3$, the data points align on straight lines in log-log scale, with slopes close to the predictions of the scaling laws developed in subsection \ref{sec:small-pe-limit} (Eqs.~\eqref{eq:SmallPelimit_delta} and  \eqref{eq:SmallPelimit_u}).

\begin{figure}
  \centering
  \includegraphics{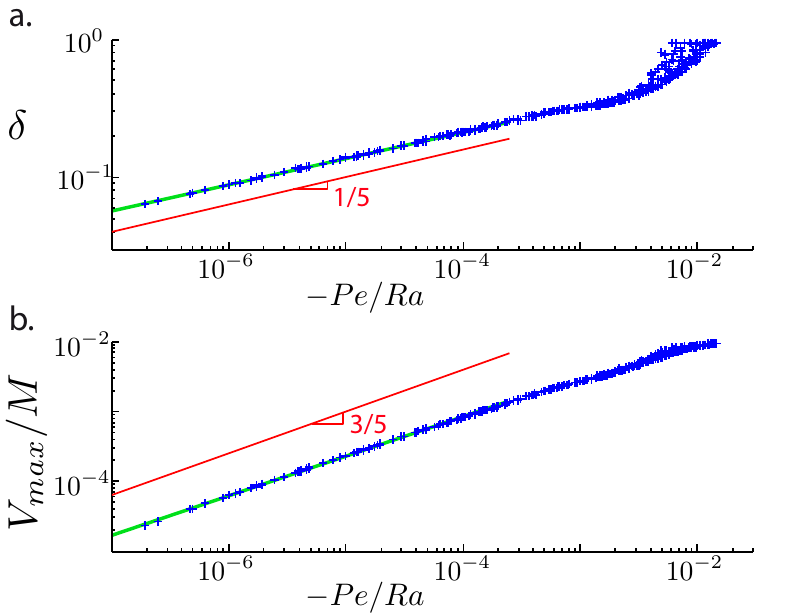}
  \caption{Thickness of the uppermost vorticity layer (a)  and maximum velocity
    (b) as functions of $-Pe/Ra$. 50 runs are plotted, with $Ra_0$ from $-3\times 10^3$ to
    $-10^{10}$ and $Pe_0$ from $13$ to 5000, with 10 time steps for each runs, and filtered
    by $Pe_0\gg 1/\delta$. Solid green lines are the best fit for $\delta <0.25$, which is
    $\delta=1.24\, (-Pe/Ra)^{0.192 \pm 0.04}$ and $V_\textrm{max}= 0.143 \,M\,
    (-Pe/Ra)^{0.56\pm0.09}$. The red lines are expected scaling laws developed in
    subsection~\ref{sec:large-pe-limit}. }
  \label{fig:delta_Vmax}
\end{figure}

Fig.~\ref{fig:delta_Vmax} shows the vorticity layer thickness and maximum velocity as functions of $Pe/(-Ra)$, in log-log scale, for runs in the large P\'{e}clet limit.
The thickness of the upper layer and the maximum velocity align on slopes close to the 1/5 and 3/5 slopes
predicted in subsection~\ref{sec:large-pe-limit}. 
Fig.  \ref{fig:delta_Vmax} has been constructed from runs with $M=1$, but we have checked that, as long as the condition $-Ra\gg M$ is verified, the geometry of the flow does not depend on $M$ and that the velocity is proportional to $M$.

\section{Strain rate} 
\label{sec:strain-rate-produced}

\begin{figure*}
  \centering
  \includegraphics{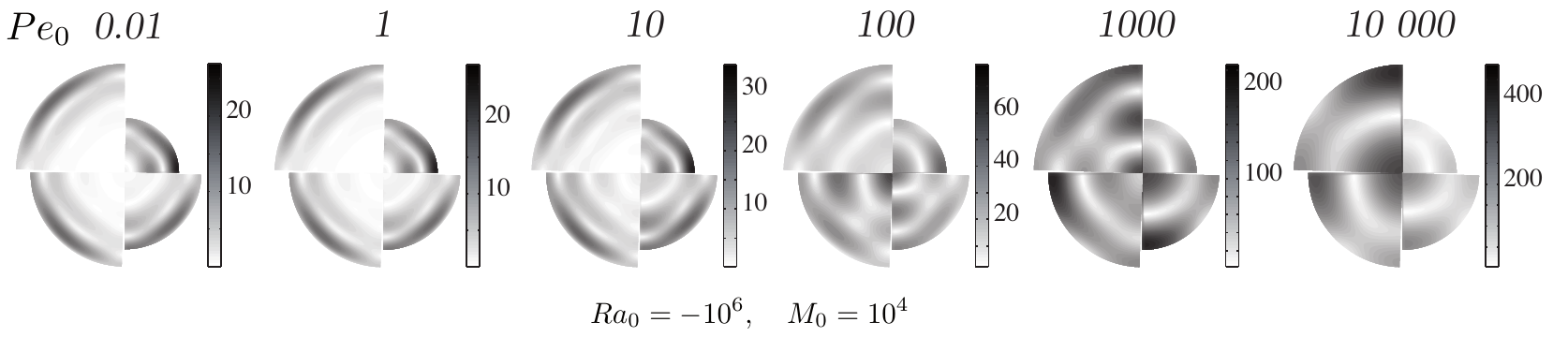}
  \caption{
  Snapshots of the von Mises equivalent strain rate for simulations with dimensionless parameters $M_{0}=10^4$, $Ra_{0}=-10^6$,   
and $Pe_{0}=0.01$, 1, 10, $10^{2}$, $10^{3}$ and $10^{4}$ (from left to right), with $r_{ic} \propto t^{1/2}$. Each panel corresponds to one simulation, with four time steps
  represented : $t=0.25$, $0.50$, $0.75$ and $1$ dimensionless time, from top-right and
  going clockwise.
 See Fig. \ref{fig:timeseries_vorticity} for plots of the vorticity field 
    of corresponding runs. }
  \label{fig:vonMises_vorticity}
\end{figure*}

 Fig.~\ref{fig:vonMises_vorticity} shows the von Mises equivalent strain
 rate  for the runs corresponding to
Fig.~\ref{fig:timeseries_vorticity}, highlighting regions of high deformation. The von Mises
equivalent strain rate is the second invariant of the strain rate tensor, measuring the power dissipated by
deformation \citep{Tome:1984ul,Wenk:2000ww,Tackley:2000vh}. Comparing
Figs~\ref{fig:timeseries_vorticity} and \ref{fig:vonMises_vorticity}, we see that the
deformation and vorticity fields have a similar geometry when the flow is organized in several
layers, 
whereas the location of the regions of high deformation and high vorticity differ when the effect of stratification is small and the flow is organized in one cell only. 
In this case, which is similar to that studied by
\citet{Karato:1999tt}, the maximum deformation is at the edges of the cells, whereas in the case of 
large stratification, the strain is confined in the uppermost layer.  
In the strongly stratified regime, the deformation can be predicted from the scaling laws
discussed in section~\ref{sec:scal-laws-thickn}, as $\dot{\epsilon}\sim u/\delta$ in dimensionless
form, or  $\dot{\epsilon}\sim u\kappa /\delta r_\mathrm{ic}^2 $ in dimensional form.

In the small $Pe$ number case, relevant for the Earth's inner core with a thermal stratification, using the scaling laws \eqref{eq:SmallPelimit_delta} and \eqref{eq:SmallPelimit_u} for the velocity and shear layer thickness gives
\begin{equation}
  \label{eq:instantStrainRate_smallPe}
  \dot{\epsilon}(t)\sim \frac{\kappa}{r_\mathrm{ic}^2(t)} \frac{u}{\delta} \simeq 0.2 \frac{\kappa}{r_\mathrm{ic}^2(t)} M(t) (-Ra(t))^{-1/3}.
\end{equation}

In the large $Pe$ number case, relevant for the Earth's inner core with a
compositional  stratification, using the scaling laws \eqref{eq:thickness_Pelarge_} and \eqref{eq:vorticity_Pelarge} for the velocity and shear layer thickness gives
\begin{equation}
  \label{eq:instantStrainRate_largePe}
  \dot{\epsilon}(t)\sim  \frac{\kappa}{r_\mathrm{ic}^2(t)} \frac{u}{\delta} \simeq 0.1 \frac{\kappa}{r_\mathrm{ic}^2(t)} M(t) \left ( \frac{-Ra(t)}{Pe(t)}\right ) ^{-2/5}.
\end{equation}
Notice that $\dot \epsilon$ is proportional to $\eta^{-2/3}$ and
$\eta^{-3/5}$ in the thermal and compositional cases, and therefore
increases with decreasing viscosity in spite of the fact that the
velocity in the shear layer decreases with decreasing $\eta$. This is
because the thickness of the shear layer decreases with viscosity
faster than the velocity.

These scaling laws give upper bounds for the actual strain rate  in the inner core, which  evolves with
time because of the time dependence of the  parameters.
The quantity $1/\dot \epsilon$ is the time needed to deform the shear layer to a cumulated strain $\sim 1$.

To estimate the cumulated deformation  in the inner core, we
assume that the strain rate $\dot{\epsilon}$ found above is applied only on the uppermost shear layer of thickness
$\delta$.
The
simplest is to assume that both $\dot{\epsilon}$ and $\delta$ are evolving slowly with
time, on a timescale long compared to the time $\delta/u_{ic}$ over which a layer of thickness $\delta$ is crystallized. 
Assuming that the strain rate $\dot{\epsilon}(t)$ is given by $u/\delta$ within the shear layer of thickness $\delta$ and is negligible elsewhere, 
the cumulated deformation is then
\begin{equation}
\epsilon \sim \frac{\dot \epsilon \delta }{u_{ic}} \sim  \frac{u}{Pe},
\label{eq:CumulatedStrain}
\end{equation}
with $u$ the dimensionless maximum horizontal velocity given by Eqs. \eqref{eq:SmallPelimit_u} or \eqref{eq:vorticity_Pelarge} depending on the value of the P\'eclet number. 
With $u$  dimensional, the cumulated strain can also be written as
\begin{equation}
\epsilon  \sim \frac{u}{u_{ic}}.
\end{equation}
The deformation magnitude below the upper shear layer is given by the ratio between the horizontal velocity
induced by the Lorentz force and the growth rate of the inner core. 

A  more elaborated method of estimating $\epsilon$ is discussed in appendix~\ref{sec:integr-over-time}. The results are close to what Eq. \eqref{eq:CumulatedStrain} predicts for $r > 0.3\, r_{ic}(\tau_{ic})$, but are more accurate for smaller $r$.
The validity of both estimates is restricted to conditions under which the strong stratification scaling laws applies, which requires that $M \ll -Ra$.
This is only verified if the inner core radius is larger than $r_{ic}(\tau_{ic}) (M_{0}/Ra_{0})^{1/4}\simeq 0.02\, r_{ic}(\tau_{ic})$ in the thermal case, and $r_{ic}(\tau_{ic}) (M_{0}/Ra_{0})^{1/5}\simeq 0.07\, r_{ic}(\tau_{ic})$ in the compositional case.

\section{Application to the inner core}
\label{sec:discussr}

To determine in which regime is  the inner core, the first step is to estimate the ratio $-Ra/M$ 
\begin{equation}
  \label{eq:Ra/M}
  \frac{-Ra}{M}=\frac{-\Delta \rho g_\mathrm{ic} r_\mathrm{ic}\mu_0}{B_0^2}=\frac{-\Delta \rho}{1\,\mathrm{kg.m}^{-3}}\left ( \frac{3\,\mathrm{mT}}{B_0}
  \right )^2 7.5\times 10^{5} .
\end{equation}
Notice that this does not depend on the viscosity. 
Plausible dimensionless numbers for the Earth's inner core are obtained from typical values
given in Table~\ref{tab:values} and summarized in Table~\ref{tab:dimensionlessparameters}. 
The density stratification is $\Delta \rho \sim 1$\,kg.m$^{-3}$ irrespectively of the nature of the stratification \citep{Deguen:2011ga,Labrosse:2014hf}.
Varying the parameters within their uncertainty range can change the ratio $-Ra/M$ by an order of magnitude at most. 
The ratio $-Ra/M$ is thus unlikely to be smaller than 1, irrespectively of the thermal or compositional origin of the
stratification. 
The inner core is strongly stratified compared to magnetic forcing. 

If the stratification is of thermal origin, the P\'{e}clet number is on the order of 1 ($Pe_0=2.8$).
Fig.~\ref{fig:timeseries_vorticity} shows that the low P\'{e}clet number scaling laws still agree reasonably well with the numerical results for P\'eclet numbers around 1; the low P\'{e}clet number scaling laws can therefore be used to predict the flow geometry and strength in the thermal stratification case.
If the stratification is of compositional origin, the P\'eclet
number is large ($Pe_0\sim 10^5$) and thus the large P\'eclet limit scaling laws apply.

Estimates of the thickness of the upper layer, of the maximum velocity in this layer and of the
expected strain rate are given in Table~\ref{tab:estimations}, using values of parameters
of Tables~\ref{tab:values} and \ref{tab:dimensionlessparameters}. Because the viscosity is
poorly known, we express these estimates as functions of the viscosity.  
As an exemple, assuming a viscosity of $10^{16}$\,Pa.s gives a shear layer thickness of 94\,km and 60\,km in case of thermal and compositional
stratification.
The velocity in this layer is expected to be  several orders of  
magnitude lower than the growth rate, and instantaneous deformation due to this flow is
 small: the typical timescale for the deformation is of order $10^2$\,Gyrs for both cases, for
$\eta=10^{16}$\,Pa.s.  These values are obtained for the present inner core, which means it
is the deformation time scale for the present uppermost layer. The deformation $\epsilon\sim
u/Pe$ is a decreasing function of the radius, and thus  is higher in depth: compared to below the ICB, the strain at
$r=0.5\,r_\mathrm{ic}$ is multiplied by 2 for thermal stratification, and by 4.6
for compositional stratification.

As can be seen in Eqs.~(\ref{eq:instantStrainRate_smallPe}) and
(\ref{eq:instantStrainRate_largePe}), the strain rate in the shear layer is a decreasing
function of the stratification strength. 
This is the opposite of what  \citet{Deguen:2011ck} found in the case of a flow forced by
heterogeneous inner core growth \citep{Yoshida:1996p54}. 
The flow geometry is similar to what has been found here if the inner core is stably
stratified, with a shear layer below the ICB in which deformation is localized, but,
contrary to the case of the Lorentz force, the strength of the flow and strain rate increase with
the strength of the stratification. 
This difference is due to the fact that the
velocity is imposed by the boundary conditions at the ICB in the
case of heterogeneous inner core growth\sout{ case}, and therefore does not decrease
when the stratification strength is increased. 
In contrast, the velocity in the shear layer produced by the Lorentz force depends on a
balance between the Lorentz force and the viscous forces, and decreases with increasing
stratification strength.

Using the scaling laws developed
above (Eqs.~\eqref{eq:instantStrainRate_smallPe}, \eqref{eq:instantStrainRate_largePe} and \eqref{eq:CumulatedStrain}), the cumulated strain below the shear layer is given by 
\begin{equation}
\epsilon_T \sim 5.6 \times 10^{-4} \, \left (
\frac{10^{16}\,\mathrm{Pa.s}}{\eta}\right ) ^{1/2} \left( \frac{B_{0}}{3\, \mathrm{mT}} \right)^{2},\label{eq:strain_thermique}
\end{equation}
and 
\begin{equation}
\epsilon_C \sim 2.4 \times 10^{-4} \,  \left (
\frac{10^{16}\,\mathrm{Pa.s}}{\eta}\right ) ^{2/5} \left( \frac{B_{0}}{3\, \mathrm{mT}} \right)^{2}.\label{eq:strain_chimique}
\end{equation}
This shows that a viscosity lower than $10^{10}$\,Pa.s is required to obtain 
 a deformation larger than about 1. Such a low viscosity seems unrealistic and this suggests
that no detectable anisotropy would be produced in the bulk of the inner core.

Using the method in appendix~\ref{sec:integr-over-time} to estimate the strain,
the  deformation is found to be two orders of magnitude larger close to the center of the
 inner core than at the edge. This means a non-negligible strain
 for viscosity lower than $10^{12}$\,Pa.s. 

 The uppermost
 layer has a different behavior because it \sout{has} does not have enough time to deform. This could
 stand for an isotropic layer at the top of the inner core, as observed by seismic
 studies. We expect this layer to be of the order of one hundred kilometers thick
 for thermal or compositional stratification.

\begin{table*}
  \centering
  \caption{Estimates of the thickness, maximal horizontal  velocity and strain rate of the upper layer
    for thermal stratification (low $Pe$) and compositional stratification (large $Pe$).}
\def\arraystretch{2.2}
  \begin{tabular}{@{}lll@{}}
      \hline
    &Thermal stratification, low P\'eclet&Compositional stratification, large P\'eclet \\
    \hline
    Thickness $\delta$ &$\left ( \dfrac{\eta}{10^{16}\,\mathrm{Pa.s}}\right) ^{1/6}\quad \times
    94$\,km&$\left ( \dfrac{\eta}{10^{16}\,\mathrm{Pa.s}}\right) ^{1/5}\quad \times 60$\,km\\
    Maximal horizontal  velocity$^a$ $u$&$\left ( \dfrac{10^{16}\,\mathrm{Pa.s}}{\eta}\right )
    ^{1/2} \quad \times  2.2\times10^{-14}\,$m.s$^{-1} $&$\left ( \dfrac{10^{16}\,\mathrm{Pa.s}}{\eta}\right )
    ^{2/5} \quad  \times 0.9\times10^{-14}\,$m.s$^{-1} $\\
    Instantaneous strain rate$^b$ $\dot \epsilon$&$\left (
      \dfrac{10^{16}\,\mathrm{Pa.s}}{\eta}\right )
    ^{2/3}\quad  \times 7.4\times 10^{-12}\,$\,yrs$^{-1}$&$\left (
      \dfrac{10^{16}\,\mathrm{Pa.s}}{\eta}\right )
    ^{3/5}\quad  \times 4.8\times 10^{-12}\,$\,yrs$^{-1}$\\
    Strain $\epsilon=u/Pe$&$\left ( \dfrac{10^{16}\,\mathrm{Pa.s}}{\eta}\right )
    ^{1/2} \dfrac{r_\mathrm{ic}}{r}\quad  \times 5.6 \times 10^{-4}$&$\left ( \dfrac{10^{16}\,\mathrm{Pa.s}}{\eta}\right )
    ^{2/5} \left ( \dfrac{r_\mathrm{ic}}{r}\right ) ^{11/5} \quad  \times 2.4\times 10^{-4}$\\
        \hline
  \end{tabular}
  \label{tab:estimations}

  \begin{flushleft}
    $^a$ this value has to be compared with a typical value for the growth rate:
   $u_{ic}(\tau_{ic})\approx 10^{-11}$m.s$^{-1}$. 

 $^b$ at $t=\tau_{ic}$.  
  \end{flushleft}

\end{table*}

\section{Conclusion and discussion}
\label{sec:concl}

Following previous studies  \citep{Karato:1999tt,Buffett:2001ul}, we have developed a complete
model for evaluating the deformation induced by the Lorentz force in a stratified inner
core, investigating the effect of boundary conditions and neutral and strong stratification
in the case of thermal or compositional stratification. 

Calculating the flow for neutral stratification with different mechanical boundary
conditions, we show that the boundary conditions depend on the
  values of the viscosity. If the viscosity is low, the ICB acts as an impermeable boundary, with no radial flow across
 the ICB, whereas if the viscosity is large  the ICB acts as a permeable boundary, with  fast melting and solidification
  at the ICB. 
  We find that 
 the velocity is larger than the inner core growth rate  if the
 viscosity is lower than $10^{16}$\,Pa.s. Unlike previous studies, the boundary conditions assumed here are of impermeable type.

If the inner core has a stable density stratification, then we find that the
stratification strongly alters the flow induced by the poloidal component of the Lorentz
force. The deformation is concentrated in a thin shear layer at the top of the inner core, which
thickness does not depend on the magnetic field strength, but depends on both the density stratification
and the P\'{e}clet number, which compares the timescales of inner core  growth and diffusion. 

However, the deformation rate in this regime is predicted to be too small for producing
significant LPO in most of the inner core, unless the inner core viscosity is smaller than
$10^{10}-10^{12}$\,Pa.s. The cumulated deformation can be two orders of magnitude larger close to the
center  of the inner core, but remains smaller than 1 if the inner core viscosity is larger than
$10^{12}$\,Pa.s.

We have made a number of simplifying assumption, but relaxing them is unlikely to significantly alter our conclusions.
Our  estimated values of the deformation are probably upper-bounds. 
The effective  strain in the inner core induced by the poloidal Lorentz force is expected to be even smaller
than these values. 
Indeed, we use assumptions that maximize the strain rate in the inner
core. 
(i) The geometry and strength of the magnetic field have been chosen to maximize
the effect of the Lorentz force: the degree (2,0) penetrates deeper in the inner core than
higher orders (smaller length-scales) and is less likely to vary with time. 
The smaller scale components of the magnetic field vary on a shorter timescale, which make them more sensitive to skin effect.
(ii) We have assumed  the magnetic field to be time-independent, a reasonable assumption because the fluctuations associated with outer core dynamics occur on a timescale short compared to the timescale of inner core dynamics.
The evolution of the magnetic field strength on the timescale of inner core growth is poorly constrained, but seems unlikely to have involved order of magnitude variations.
(iii) Though a growth law of the form $r_\mathrm{ic}\propto t^{1/2}$ has been assumed in the numerical simulations, our derivation of the scaling laws (section \ref{sec:scal-laws-thickn}) makes no assumption on the inner core growth law, and our scaling law should therefore also apply if this assumption is relaxed.

Finally, we have focused on the effect of the azimuthal component of the magnetic field, which produces a poloidal Lorentz force in the inner core, and have left aside the combined effect of the azimuthal and $z-$component (parallel to Earth's spin axis), which produce an azimuthal Lorentz force driving an azimuthal flow \citep{Buffett:2001ul}.
There is no loss of generality involved, because the axisymmetric poloidal flow we have investigated and the azimuthal flow calculated by \citet{Buffett:2001ul} are perfectly decoupled, and add up linearly. 
The azimuthal flow is horizontal, and is therefore not affected by the thermal and compositional fields and their perturbations by the axisymmetric poloidal flow. Conversely, since the flow and density perturbations induced by the axisymmetric poloidal Lorentz force are axisymmetric, they are not affected by an azimuthal flow.
The azimuthal flow velocity is 
\begin{equation}
v_{\phi} = - \frac{1}{10} \frac{B_{z} B_{\phi} }{\mu_0\, \eta} \frac{r^{3}}{r_\mathrm{ic}^{2}} \sin \theta
\end{equation}
\citep{Buffett:2001ul}, and the associated strain rate is
\begin{equation}
\dot \epsilon (r,\theta) = \dot \epsilon_{r,\phi} (r,\theta)  = \frac{1}{10}\frac{B_{z}B_{\phi}r^{2}}{\mu_{0} \eta r_\mathrm{ic}^{2}} \sin \theta. 
\end{equation}
This is always larger than the strain rate predicted by our scaling laws in the strongly stratified regime (Eqs. \eqref{eq:instantStrainRate_smallPe} and \eqref{eq:instantStrainRate_largePe} for thermal and compositional stratification), which implies that this deformation field will dominate over deformation due to the poloidal component of the Lorentz force.

\section*{Acknowledgments}

We would like to thank two anonymous reviewers for their constructive comments, and Christie Juan for her help at a preliminary stage of this project.
RD gratefully acknowledges support from grant ANR-12-PDOC-0015-01 of the ANR (Agence Nationale de la Recherche). SL was supported by the Institut Universitaire de France for this work.

\bibliographystyle{gji}

\appendix

\section{Poloidal/toroidal decomposition}
\label{sec:poloidal-form-set}

\subsection{Momentum equation using poloidal decomposition}
\label{sec:moment-equat-using}

Following \citet{Ricard:1989vf} and \citet{RIBE-TG2007}, a flow induced by internal density anomalies in a
constant viscosity fluid is purely poloidal in a spherical shell, if the surface
boundary conditions also have a zero vertical vorticity. Considering also that the
external forcing by the Lorentz force is purely poloidal ($\mitbf{r}\cdot  \mitbf{\nabla}\times
\mitbf{F}_L=0 $),  no toroidal flow is expected in
our problem. The velocity field can thus  be written by introducing the  scalar $P$
defined 
such that
\begin{equation}
  \label{eq:poloidalvelocityfield}
  \mitbf{u}=\mitbf{\nabla}\times \mitbf{\nabla} \times (P\mitbf{r}),
\end{equation}
where $\mitbf{r}=r\mitbf{e}_r$ is the position vector. 

Applying $\mitbf{r}\cdot (\mitbf{\nabla}\times \mitbf{\nabla}\times\,\, )$ to Eq.~(\ref{eq:dimensionlessSetofEquations2}) , we
obtain that
\begin{equation}
  \label{eq:poloidal_momequation}
  0=Ra(t)L^2\Theta - (\nabla ^2) ^2 L^2 P+M(t)\mitbf{r}\cdot (\mitbf{\nabla}\times \mitbf{\nabla}\times \mitbf{F}_L), 
\end{equation}
 where $L^2$ is the operator defined by
 \begin{equation}
   \label{eq:L2}
   L^2=-\frac{1}{\sin \theta}\derpar{}{\theta} \left ( \sin \theta \derpar{}{\theta}\right )
 - \frac{1}{\sin^2 \theta } \derpar{^2}{\phi ^2}. 
 \end{equation}

The last term of Eq.~(\ref{eq:poloidal_momequation}) is computed from the Lorentz
force defined in Eq.~(\ref{eq:fL}) and gives
\begin{equation}
  \label{eq:Lorentzforce_doublerot}
  \mitbf{r}\cdot  (\mitbf{\nabla}\times \mitbf{\nabla}\times \mitbf{F}_L)=8  r^2 (1-3\cos ^2 \theta)=  -\frac{16}{\sqrt{5}}r^2 Y_2^0,
\end{equation}
where $Y_2^0=\frac{\sqrt{5}}{2}(3\cos ^2 \theta -1)$.

Eq.~(\ref{eq:poloidal_momequation}) becomes 
\begin{equation}
  \label{eq:poloidal_momequation_2}
   0=Ra(t)L^2\Theta - (\nabla^2)  ^2 L^2 P-M(t)\frac{16}{\sqrt{5}}r^2 Y_2^0.
\end{equation}

When expanding the two scalar field $\Theta$ and $P$ with spherical harmonics $Y_l^m(\theta,\phi)$
that satisfy $L^2Y_l^m=l(l+1)Y_l^m$, we defined new variables $t_l^m$ and $p_l^m$ by
\begin{eqnarray}
  \Theta&=&t_l^m(r)Y_l^m,\\
  P&=&p_l^m(r)Y_l^m, \,\,\, l \geq  1.
\end{eqnarray}

Eq.~(\ref{eq:poloidal_momequation_2}) is eventually written as 
\begin{equation}
  \label{eq:poloidal_momequation_3}
  D_l^2p_l^m+\frac{16}{\sqrt{5}l(l+1)}M(t)r^2\delta_{2l}\delta_{0m}-Ra(t)t_l^m=0,
  \,\,\, l\geq  1,
\end{equation}
where $\delta$ is the Kronecker symbol and $D_{l}$ is the second order differential operator defined by
\begin{equation}
  \label{eq:Dl}
  D_{l}=\frac{d^2}{dr^2}+\frac{2}{r}\frac{d}{dr}-\frac{l(l+1)}{r^2}.
\end{equation}

\subsection{Poloidal decomposition of the boundary conditions}
\label{sec:polo-decomp-bound}

From \citet{Deguen:2013bj}, the boundary conditions at $r=1$ are written as 
\begin{eqnarray}
  \tau_{r\theta}=\eta\left [ r\derpar{}{r}\left (\frac{u_{\theta}}{r} \right )
    +\frac{1}{r}\derpar{u_r}{\theta}\right ]&=0,  \label{eq:BC_total-1}\\
  \tau_{r\phi}=\eta\left [ r\derpar{}{r}\left (\frac{u_{\phi}}{r} \right )
    +\frac{1}{r\sin \theta}\derpar{u_r}{\phi}\right ]&=0,  \label{eq:BC_total-2}\\
  -\mathcal{P}(t)(u_r-u_{ic})-2\derpar{u_r}{r}+p'&=0.  \label{eq:BC_total-3}
\end{eqnarray}
$\mathcal{P}(t)$ is the dimensionless parameter that characterizes the resistance to phase
change as defined in Eq.~(\ref{eq:Pt}). 

In term of poloidal decomposition of the velocity field, the set of equations for the
boundary conditions at $r=1$ is modified as
\begin{equation}
  \frac{d^2p_l^m}{dr^2}+[l(l+1)-2]\frac{p_l^m}{r^2}=0, \, \,\, l\geq
  1,\label{eq:BC_total-1_poloidal}
\end{equation}
\begin{equation}
  r\frac{d^3p_l^m}{dr^3}-3l(l+1)\frac{1}{r}\frac{dp_l^m}{dr}=\left [
    l(l+1)\mathcal{P}(t)-\frac{6}{r^2}\right ]p_l^m, \,\,\, l\geq1.\label{eq:BC_total-2_poloidal}
\end{equation}


\section{Thermal stratification}
\label{sec:therm-strat}

We derive here the expression of the reference diffusive potential temperature profile given in Eq. \eqref{eq:Theta_r},  under the assumption of $r_{ic} \propto t^{1/2}$.

The reference potential temperature is the solution of
\begin{equation}
  \derpar{\Theta}{t}-\tilde{r}\frac{u_{ic}(t)}{r_{ic}(t)} \frac{\partial \Theta}{\partial r} =\frac{\kappa_{T}}{r^2_{ic}(t)} \frac{1}{\tilde{r}^2} \frac{\partial}{\partial r}\left( \tilde{r}^{2} \frac{\partial \Theta}{\partial \tilde{r}}\right) +S_{T}, 
  \label{eq:HeatAppendix}
\end{equation}
obtained from  Eq. \eqref{eq:temperature_r'x} by taking $\chi=\Theta$ and $\mitbf{u}=\mitbf{0}$.
The source term $S_{T}$ is constant if $r_{ic}\propto t^{1/2}$.
The potential temperature $\Theta$ is a function of $\tilde{r}$, $\kappa_{T}$, $r_{ic}(t)$, $S_{T}$ and $u_{ic}(t)$ only. According to the Vaschy-Buckingham theorem, we can form only three independent dimensionless groups  from these variables (6 dimensional variables  - 3 independent physical units), one possible set being $\Theta/(S_{T}r_{ic}^{2}/\kappa_{T})$, $u_{ic}r_{ic}/\kappa_{T}$, and $\tilde{r}$. With $r_{ic} \propto t^{1/2}$, $u_{ic}r_{ic}/\kappa_{T}$ is constant and equal to $Pe_{0}$.
The potential temperature must therefore be of the form
\begin{equation}
\Theta = \frac{S_{T} r_{ic}^{2}(t)}{\kappa_{T}} f(\tilde{r},Pe_{T0}).
\label{eq:ThetaScaling}
\end{equation}
By definition, the potential temperature is equal to 0 at the ICB, which implies $f(\tilde{r}=1,Pe_{T0})=0$. 
With $r_{ic}(t) = r_{ic}(\tau_{ic}) (t/\tau_{ic})^{1/2}$ and noting that $r_{ic}^{2}(\tau_{ic})/(\kappa_{T} \tau_{ic}) = \xi_{T0}=2 Pe_{T0}$ (see Eqs. \eqref{eq:parameters_xi} and \eqref{eq:parameters_functionoft_xi}), inserting Eq. \eqref{eq:ThetaScaling} into Eq. \eqref{eq:HeatAppendix} yields
\begin{equation}
0 = f''+ f' \left(\frac{2}{\tilde{r}}+Pe_{T0} \tilde{r}\right) - 2Pe_{T0} f +1,
\end{equation}
where $f'$ and $f''$ stand for the first and second derivatives of $f$ with respect to $\tilde{r}$.
Looking for a polynomial solution in $\tilde{r}$ satisfying $f(\tilde{r}=1,Pe_{T0})=0$, we find that $f =(1-\tilde{r}^{2})/(6+2Pe_{T0})$, which gives
\begin{equation}
\Theta = \frac{S_{T} r_{ic}^2(t)}{6\kappa_{T} (1+Pe_{T0}/3)}
\left[1-\left(\frac{r}{r_{ic}(t)}\right)^2\right]	\label{eq:Theta_r_2}.
\end{equation}


\section{Compositional stratification}
\label{sec:comp-strat}

The source term $S\!_c$ of the conservation of light elements is directly related to the evolution of
the concentration of light elements in the solid that freezes at the inner core
boundary $\dot c_{ic}^s(t)$. Following \citet{Gubbins:2013ip} and \citet{Labrosse:2014hf},
this term depends both on the evolution of the concentration in the liquid outer core,
which increases when the inner core grows because the solid incorporates less light
elements than is present in the outer core, and on the evolution of the partition coefficient between solid and liquid. 

In this paper, we will focus on the simplest case for which the partition coefficient does
not depend on temperature or concentration. 
Thus, the concentration in the solid is increasing
with the radius of the inner core, as the concentration in the liquid increases. This will
promote a  stably stratified inner core, whereas
\citet{Gubbins:2013ip} and \citet{Labrosse:2014hf} focused on  the potentially destabilizing effects of
variations of the partition coefficient.

To estimate the light element concentration, we note $M_{c}=M_{ic}+M_{oc}$ the total mass
of the Earth's core. When increasing the inner core mass by $d\, M_{ic}$, the mass of the
outer core light elements decreases by $d\,(c^lM_{oc})=-c^s_{ic}d\, M_{ic}$ . The total
mass of the Earth's core is constant, which gives $d\,M_{oc}=-d\,M_{ic}$ and
\begin{equation}
  \label{eq:composition_1}
  \frac{d\,c^l}{c^l}=(k-1)\frac{d\,M_{oc}}{M_{oc},}
\end{equation}
where $k$ is the partition coefficient defined as $k=c^s_{ic}/c^l$.

Eq.~(\ref{eq:composition_1}) can be integrated with the assumption of a  constant  partition
coefficient. Integration between $(c_0^l, M_{c})$ and $(c^l,M_{oc})$, corresponding to
before the inner core formation and any time after, this gives
\begin{equation}
  \label{eq:composition_2}
  c^l(t)=c^l_0\left ( \frac{M_{oc}}{M_c}\right )^{k-1}.
\end{equation}

When ignoring radial density variations in the outer core, the ratio $M_{oc}/M_c$ is simply
$1-(r_{ic}/r_c)^3$. 
Taking into account compressibility (radial density variations in the core) and the density jump at the ICB results in a stratification approximately 15\% larger.
The light element concentration at the inner core boundary is thus directly obtained from the liquid concentration as 
\begin{equation}
  \label{eq:composition_3}
c^s_{ic}(t)= kc^l_0\left ( 1-\left ( \frac{r_{ic}(t)}{r_c}\right )^3\right )^{k-1} . 
\end{equation}


\section{Analytic solution for neutral stratification}
\label{sec:analyt-solut-ra=0}

We solve Eq.~(\ref{eq:poloidal_momequation_3}) for a neutral stratification, $Ra=0$, with the boundary
conditions \eqref{eq:BC_total-1_poloidal} and \eqref{eq:BC_total-2_poloidal}  described  in section~(\ref{sec:polo-decomp-bound}). 

Equation
(\ref{eq:poloidal_momequation_3}) is thus written as
\begin{equation}
  \label{eq:neutralstratification_momentum}
 D_l^2p_l^m+\frac{16}{\sqrt{5}l(l+1)}Mr^2\delta_{2l}\delta_{0m}=0,  
\end{equation}
and can be solved analytically. 

Except for $(l=2,m=0)$, $p_l^m=0$ is solution of the Eq.~(\ref{eq:neutralstratification_momentum}) and verifies the boundary conditions
(\ref{eq:BC_total-1_poloidal}) and (\ref{eq:BC_total-2_poloidal}). 

For $(l=2,m=0)$, we have
\begin{equation}
  \label{eq:neutralstratification_momentum-l2m0}
 D_l^2p_2^0+\frac{8}{3\sqrt{5}}Mr^2=0 , 
\end{equation}
and for $r=1$
\begin{eqnarray}
  \frac{d^2p_2^0}{dr^2}+4\frac{p_2^0}{r^2}&=0,\label{eq:BC-1}\\
  r\frac{d^3p_2^0}{dr^3}-18\frac{1}{r}\frac{dp_2^0}{dr}&=\left[
    \mathcal{P}(t)-\frac{1}{r^2}\right]6 p_2^0.\label{eq:BC-2}
\end{eqnarray}

Eq.~(\ref{eq:neutralstratification_momentum-l2m0}) is solved considering a sum of
polynomial functions, and adding the boundary conditions (\ref{eq:BC-1}) and
(\ref{eq:BC-2}), we obtain the coefficient $p_2^0$ as
\begin{multline}
  \label{eq:p20}
  p_2^0(r)=\frac{M}{3^37\sqrt{5}}\\
\left ( -r^6+\frac{14}{5}r^4-\frac{9}{5}r^2+\frac{204}{5}\frac{r^4}{19+5P}-\frac{544}{5}\frac{r^2}{19+5P} \right ).
\end{multline}

From the coefficients $p_l^m$, the vertical and horizontal velocities are  
\begin{equation}
  \label{eq:verticalvelocity}
  u_r=\sum_{l,m}l(l+1)\frac{p_l^m}{r}Y_l^m,
\end{equation}
\begin{equation}
  \label{eq:horizontalvelocity}
  u_{\theta}=\sum_{l,m}\frac{1}{r} \frac{d}{dr}\left ( r p_l^m\right )\derpar{}{\theta}Y_l^m,
\end{equation}
with $Y_l^m$ the surface spherical harmonics. 

The root mean square velocity ($V_{rms}$) of the system is defined as 
\begin{equation}
  \label{eq:Vrms}
  V_{rms}^2=\frac{3}{4\pi}\int _0^{2\phi}\int_0^{\pi}\int_0^1 (u_r^2+u_{\theta}^2)\sin \theta r^2\, dr \, d\theta\,d\phi.
\end{equation}

From Eq.~(\ref{eq:horizontalvelocity}) and (\ref{eq:Vrms}), we obtain the maximum
absolute value of the horizontal velocity and the RMS velocity that are shown on Fig.~\ref{fig:analyticsolution}.
Both graphs have a sigmoid shape, and thus we are mostly interested in the extreme values
for each velocity, which are given in Table~\ref{tab:velocities}.

\begin{table}
  \caption{Extreme values for RMS velocity and maximum of the absolute value of the
    horizontal velocity for two extreme values of $\mathcal{P}$. Velocities are
    proportional to $M$ and thus only the value $v/M$ is given.}
  \centering  
 \label{tab:velocities}
  \begin{tabular}{l|cc}
\hline
    & $\mathcal{P}\to 0$ & $\mathcal{P}\to \infty$\\
\hline
    $V_{rms}/M$& 0.06609 & 0.00805\\
\hline
    $\max |u_{\theta}(r,\theta) |/M$ & 0.06944&0.01270
  \end{tabular}

\end{table}


\section{Integration over time of the deformation}
\label{sec:integr-over-time}

\subsection{General discussion}
\label{sec:general-discussion}

In general, the texturation mechanism is a nonlinear process, but  an upper bound of the
total deformation  can be inferred by considering  that the strain adds up linearly. 
 The material is deformed at a strain rate
$\dot{\epsilon}$ during the time $\delta/u_{ic}$ needed to grow a layer of thickness $\delta$, and so 
\begin{equation}
  \label{eq:strain_uPe}
  \epsilon(r(t))=\dot{\epsilon}(t)\frac{\delta(t)}{u_{ic}(t)}=\frac{u}{Pe},
\end{equation}
with $Pe=u_{ic}r_{ic}/\kappa$. 

This equation leads to simple forms at the low and large P\'{e}clet limits, with $\epsilon \propto t^{-1/2}$
for thermal stratification and low P\'{e}clet, and $\epsilon \propto t^{-11/10}$ for compositional
stratification and large P\'eclet. 
In what follows, we will compare the simple estimate given above with results of more elaborate calculations.

The total deformation of a given material
during a time $\tau$ can be inferred more precisely by $\int _0^{\tau} \dot{\epsilon}(t)dt$.
Using dimensionless quantities described in the main sections, 
the deformation of a
stratified sphere
subject to a magnetic forcing is 
\begin{equation}
  \label{eq:strain}
  \epsilon(r)=\int _0^1 \dot{\epsilon}(r,t)\,dt,
\end{equation}
with $\dot{\epsilon}$ the strain rate function that will be described by a rectangular function as

\begin{equation}
  \label{eq:strainrate}
  \dot{\epsilon}(r,t)=\left \{ 
    \begin{array}{ll}
\dot{\epsilon}_{vM}(t)\frac{\kappa}{r_{ic}^2(t)}\, &\text{ if } \,r_{ic}(t)(1-\delta)<r<r_{ic}(t)\\
   0 &\text{ elsewhere}.
\end{array}
\right .
\end{equation}

The estimations of $\dot{\epsilon}_{vM}$ depend on the scaling laws defined in section
\ref{sec:strain-rate-produced}, and also on the time dependence of the parameters we have
defined. 

Because $r_{ic}(t)(1-\delta)<r_{ic}(t) \,\, \forall t$, integrating over time the function
defined by (\ref{eq:strainrate}) is equivalent to integrate it between $t_{min}(r)$ and
$t_{max}(r)$, where   $t_{min}$ and
$t_{max}$ are defined by $r_{ic}(t_{max})(1-\delta)=r$ and $r_{ic}(t_{min})=r$,
\begin{equation}
  \label{eq:strain_r}
  \epsilon (r)=\int _{t_{min}}^{t_{max}}\dot{\epsilon}_{vM}(t)\frac{\kappa}{r_{ic}^2(t)}\, dt.
\end{equation}

\subsection{Low Pe - Thermal stratification}
\label{sec:lowPe}
In the low P\'{e}clet limit, the dimensionless thickness, maximal horizontal velocity and strain rate of
the uppermost layer are given by 
\begin{eqnarray}
  \delta  \sim  & (-Ra)^{-1/6},  \label{eq:SmallPelimit_delta-annexe}\\
u \sim & M \, (-Ra)^{-1/2},   \label{eq:SmallPelimit_u-annexe}\\
\dot{\epsilon}\sim& M (-Ra)^{-1/3}. \label{eq:SmallPelimit_e-annexe}
\end{eqnarray}
 
In dimensional form, $\delta$ happens to be  constant with time for thermal stratification
\begin{equation}
  \label{eq:delta_dimensional_thermal}
  \delta=1.9643r_{ic}(\tau_{ic})(-Ra_0)^{-1/6}.
\end{equation}

Thus, $t_{min}$ and $t_{max}$ are easy to defined as 
\begin{equation}
  \label{eq:tmin}
  t_{min}(r)=\tau_{ic}\left ( \frac{r}{r_{ic}(\tau_{ic})}\right ) ^2,
\end{equation}
\begin{equation}
  \label{eq:tmax}
    t_{max}(r)=\tau_{ic}\left ( \frac{r+\delta}{r_{ic}(\tau_{ic})}\right ) ^2,
\end{equation}
except for time close to $\tau_{ic}$ because the inner core has not enough time to be deformed,
and $t_{max}=\tau_{ic}$. For small radius, the limit will be defined by $\delta=1$, which is
here $-Ra(t)=M(t)$, about 28km for typical values of the parameters.

For thermal stratification and time dependence as defined previously, the instantaneous
deformation is 
\begin{eqnarray}
  \label{eq:instantdeformation}
  \dot{\epsilon}(t)\frac{\kappa}{r_{ic}(t)}&=&0.2148\,M_0(-Ra_0)^{-1/3}\frac{\kappa}{r_{ic}^2(\tau_{ic})}\tau_{ic}t^{-1},\\
    &=&\dot{\epsilon}_0 \, \tau_{ic}\, t^{-1},
\end{eqnarray}
with $\dot{\epsilon}_0$ in s$^{-1}$ and $t$ in s.

\begin{eqnarray}
  \label{eq:strain_r_2}
  \epsilon(r) = \left \{ 
  \begin{array}{ll}
\dot{\epsilon}_0\tau_{ic}\, 2\, \ln \frac{r-\delta}{r}, \, &\text{ for } r<r_{ic}(\tau_{ic})-\delta,\\
  \dot{\epsilon}_0\tau_{ic}\, 2\, \ln \frac{r_{ic}(\tau_{ic})}{r}, \, &\text{ for } r>r_{ic}(\tau_{ic})-\delta,
\end{array}
\right .
\end{eqnarray}

 The strain rate  is assumed to be
constant over the layer $\delta$ 
whereas we could have used numerical results of the simulations to have the exact
repartition and profile of strain over radius and time. But because the von Mises strain rate profile
over radius is close to a rectangular function, it is easier to work
with an analytic solution
such as the one discussed here.  It implies a linear increase of the absolute value of the
strain, which is 
unlikely. This will in general overestimate the total strain. 

\begin{figure}
  \centering
  \includegraphics{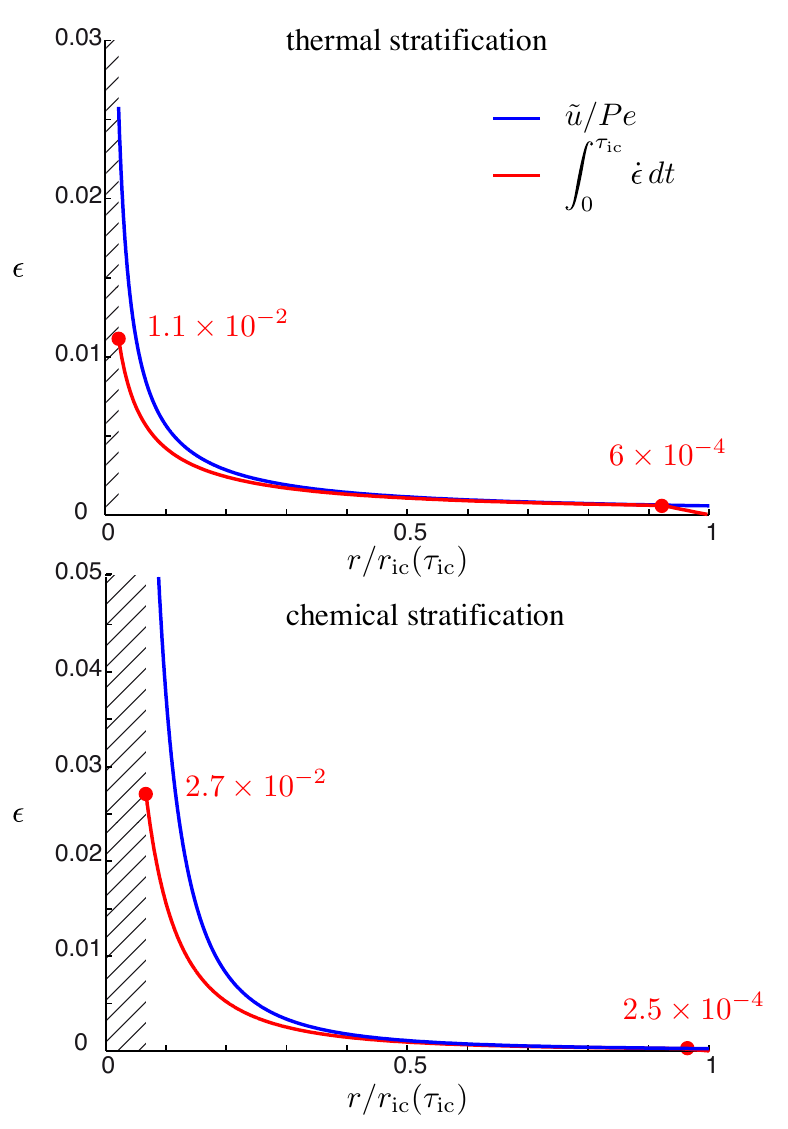}
  \caption{Strain as a function of the radius of the inner core for thermal stratification
    (a.) and compositional stratification (b.). Integration from equation
    \eqref{eq:strain_r} is the red line (analytical solution for thermal stratification
    and numerical solution for compositional stratification), and the blue lines stand for the estimation
    $u/Pe$, which is valid for $r>0.5\,r_\mathrm{ic}$. The minimum radius is computed for
  $Ra/M=1$, limit under which the strong stratification approximation is no longer
  valid.}
  \label{fig:strain}
\end{figure}

Comparison between the strain computed from (\ref{eq:strain_r_2}) and the simplest solution
$u/Pe$ discussed in the text is plotted on Fig.~\ref{fig:strain}. The solution $u/Pe$ is a
good approximation for radius larger than $0.3 \, \tau_{ic}$, except in the uppermost layer,
in which the deformation did not occur completely yet. 
It is interesting to notice that this magnetic
forcing is expected to be several orders of magnitude  more efficient when the inner core was younger.

\subsection{Large Pe - Compositional stratification}
\label{sec:large-pe-chemical}

For compositional stratification, 
\begin{equation}
  \label{eq:delta_dimensional_chemical}
  \delta=\left ( \frac{Pe}{-Ra_0}\right ) ^{1/5}r_{ic}(\tau_{ic}) \left (
    \frac{t}{\tau_{ic}}\right ) ^{-1/5}.
\end{equation}

No exact solution for inverting
$r_{ic}(\tau_{ic})(t_{max}/\tau_{ic})^{1/2}-\delta(\tau_{ic})(t/\tau_{ic})^{-1/5}=r$ can be
found.
Fig. \ref{fig:strain} b. shows the strain rate according to  numerical integration and the
approximation 
$u/Pe$. $u/Pe$ is a good approximation for $r>0.3\, r_\mathrm{ic}$.

\end{document}